\documentclass[lettersize,journal]{IEEEtran}
\usepackage{amsmath,amssymb,amsfonts}
\usepackage{algorithmic}
\usepackage{algorithm}
\usepackage{array}
\usepackage[caption=false,font=normalsize,labelfont=sf,textfont=sf]{subfig}
\usepackage{textcomp}
\usepackage{stfloats}
\usepackage{url}
\usepackage{verbatim}
\usepackage{graphicx}
\usepackage[colorlinks=true,linkcolor=black,anchorcolor=black,citecolor=black,filecolor=black,menucolor=black,runcolor=black,urlcolor=black]{hyperref}
\usepackage{color}
\usepackage{scalefnt}
\usepackage{cite}
\begin{document} 
\scalefont{1.05}
\title{\huge{Pilot-Aided Simultaneous Communication And Localisation (PASCAL) Under Practical Imperfections}}

\author{Shuaishuai Han,~\IEEEmembership{Member,~IEEE,} Mohammad Al-Jarrah,~\IEEEmembership{Member,~IEEE,} Emad Alsusa, 
\IEEEmembership{Senior
Member,~IEEE}
\thanks
{Shuaishuai Han, M. A. Al-Jarrah, and E. Alsusa are with the Department of Electrical and Electronic Engineering,
University of Manchester, Manchester M13 9PL, U.K. (e-mail: {shuaishuai.han@postgrad.manchester.ac.uk,}
\{mohammad.al-jarrah, e.alsusa\}@manchester.ac.uk).}\thanks
}

\date{}
\maketitle

\begin{abstract}
This paper introduces a system model called pilot-aided simultaneous communication and localisation (PASCAL) and illustrates its performance in the presence of practical gain and phase imperfections. Specifically, we consider the scenario where multiple single-antenna unmanned aerial vehicles (UAVs) transmit data packets to a multi-antenna base station (BS) that has the dual responsibility of detecting communication signals and localising UAVs using their pilot symbols. Two forms of receiver signal processing approaches are adopted, including disjoint localisation and communication by using maximum likelihood estimation, as well as joint localisation and data detection achieved by the newly proposed algorithms. To evaluate the asymptotic localisation performance in the presence of gain-phase imperfections, the average Cramér-Rao lower bound (CRLB) is derived, while for evaluating the communication's performance, the average sum data rate (SDR) for all the UAVs is derived in closed-form. It is shown that these derived expressions concur with simulations. The results reveal that while the proposed PASCAL system can be sensitive to gain-phase imperfections, it remains to be a powerful and efficient means to achieve reliable simultaneous localisation and communications.
\end{abstract}

\begin{IEEEkeywords}
 Average sum data rate (SDR), Average Cramér-Rao lower bound (CRLB), maximum likelihood, pilot-aided simultaneous communication and localisation (PASCAL)
\end{IEEEkeywords}

\maketitle

\section{INTRODUCTION}
In the past decades, localisation and communication used to operate in parallel using different types of equipment and network resources. However, this traditionally separated design underutilizes the limited resources. Therefore, integrated sensing and communication (ISAC) systems have been sought \cite{ISACJarrah1}-\cite{OJ-COMS3}. Compared to the separated localisation and communication systems, the ISAC model allows both systems to share the same network resources and hardware appliances \cite{OJ-COMS4}. Even if ISAC is a promising model, most of the existing literature usually ignores the fact that targets to be localised have transceiver units, and most fall in the category of cooperative users. Thus, in this work, we aim to use this kind of users and leverage their transceiver units to actively send signals to a base station (BS) not only for communicating with the BS but also for localisation by utilising the existing pilot symbols in a data frame. It is noteworthy highlighting that the introduced pilot-aided simultaneous communication and localisation (PASCAL) saves significant energy as the signal received at the BS suffers from one-way path loss rather than two-way path loss experienced in conventional radar-like localisation or the localisation conducted in the ISAC model, which typically emits signals toward targets and performs the localisation based on echos. PASCAL can be applied to many applications including, but not limited to, unmanned aerial vehicles (UAVs), vehicle-to-everything (V2X) \cite{OJ-COMS5} and Internet of things (IOT) devices,  and many location-aware communication services \cite{location-aware communication}. 

To boost communication and localisation performance, high-performance localisation algorithms and efficient communication signal pre-processing techniques are indispensable. Efficient localisation algorithms include the estimation of signal parameters via rotational invariance techniques (ESPRIT), multiple signal classification (MUSIC) \cite{MUSIC1} and maximum likelihood estimator (MLE). Both ESPRIT and MUSIC are parameter estimation algorithms, where the former is search-free and provides point estimates, while the latter estimates source parameters by searching for the peaks in the spatial spectrum. Compared to ESPRIT and MUSIC, the MLE provides the optimal localisation performance. However, these algorithms rely on the assumption of ideal phase synchronization for the array elements, which may not be guaranteed in real life. Thus, \cite{Non-Coherent estimation} considers the non-coherent estimation that is suitable for imperfect phase synchronization. On the other hand, signal pre-processing methods for communications include maximum ratio combining (MRC) and minimum mean square error (MMSE), in which MRC has the optimum performance for independent additive white Gaussian noise channels, where multiple-access interference (MAI) is absent. When MAI exits, MMSE is preferable as MRC lacks optimality under such conditions. However, implementing MMSE for systems with multiple users vastly increases the computational cost compared to MRC \cite{MMSEMRC}, and thus MRC is employed here. 

\subsection{Literature Review}

In recent years, much research has been conducted towards achieving higher communication and localisation performance using the same platform. In \cite{ISACzhen}, the ISAC technique is applied to vehicle-to-infrastructure (V2I) or V2X systems, in which a roadside unit (RSU) composed of a communication unit and a radar unit is employed for achieving communication and localisation on the same platform. In specific,  the communication unit is responsible for conducting the communication functions including information transmission and reception, while the radar unit is employed to receive the echo signals to achieve localisation functions. It is worth mentioning that the estimation of location parameters is achieved by using pilot signals in \cite{ISACzhen}, which is similar to the method in \cite{ISACpan}. Due to the fact that pilots can be utilised to obtain the location parameters prior to communicating with users, the pilot optimisation is investigated in \cite{ISACzhu} and \cite{ISAChua} to determine the optimal pilot design in ISAC systems.  In \cite{ISACyuan}, the estimated location parameters are employed to compensate for the line-of-sight (LOS) channel effect prior to communicating with the users since the LOS channel can be constructed by using location parameters. However, the effect of estimation errors for location parameters is ignored. In \cite{ISACLyu}, the UAVs are integrated within the ISAC setup, where UAVs have been deployed as aerial dual-functional access points to improve the communication and localisation performance by exploiting the UAV maneuver control and strong LOS air-to-ground (A2G) links. In \cite{ISACren}, the performance tradeoff between the information transmission and target sensing is investigated in a multi-antenna ISAC system, in which Cramér-Rao bound (CRB) and achievable rate are respectively used to evaluate the communication and sensing performance. In \cite{analysis_li}, the trade-off between the probability of coverage and the probability of detection is explored under power or bandwidth constraints. Such trade-off analysis is beneficial for understanding the competitive interaction between sensing and communication functionalities, while providing guidance for developing efficient algorithms or techniques for ISAC. Nevertheless, the system model employed in \cite{ISACzhen}-\cite{analysis_li} performs localisation by using echo signals, which is less efficient compared to using pilots in the uplink data frame to achieve the same purpose in PASCAL since two-way path loss is experienced in such ISAC systems \cite{ISACzhen}-\cite{analysis_li}, while PASCAL suffers from one-way path loss. In addition, the targets to be localised and users that need to communicate belong to different services in \cite{ISACzhen}-\cite{analysis_li}, which is different from the case of PASCAL, where the targets to be localised are also the communication object in the latter system.

Albeit, neither the communication performance nor the positioning performance of the above ISAC systems can be guaranteed in real-life applications where array model errors, such as gain and phase defects, exist. The gain-phase defects are introduced due to imperfect amplifier and phase synchronization. More specifically, they are mainly caused by the phase noise of the local oscillator, the mismatch of the receiver electronic circuit, and defects in down-sampling process due to clock drifting by the local oscillator \cite{GPerrors}. Gain-phase defects are very critical and will cause erroneous localisation and communication degradation of integrated localisation-communication systems if overlooked.  Therefore, much effort has been devoted in the past decade to quantify the effect of array model defects on antenna systems. In \cite{Performance 1}, the authors provide performance evaluation using the Cramér–Rao bound (CRB) of the coarray-based MUSIC algorithm under the effect of small sensor location errors. Their results indicate that sensor position defects introduce a constant bias, which cannot be eradicated by only increasing the signal-to-noise ratio (SNR). In \cite{Performance 3}- \cite{MCRB}, CRBs or the misspecified CRB conditioned on gain-phase errors are derived, where they consider gain-phase errors as deterministic variables in deriving CRBs. Nevertheless, since gain-phase errors vary over time in real-world scenarios, it is inappropriate to assume that gain-phase errors are deterministic in the derivation of CRBs. Gain-phase errors are usually modelled by using the real Gaussian distribution \cite{Performance 3}, \cite{Performance 5} or the uniform distribution \cite{Performance 4}. However, both real Gaussian distribution and uniform distribution are oversimplified, which do not capture the characteristics of the gain-phase errors. In addition, the aforementioned works in \cite{Performance 1}-\cite{Performance 4} have considered Multiple-input-multiple-output (MIMO) radar systems only rather than integrated or simultaneous communication-localisation.

\begin{figure}[ptb]
\centering
\includegraphics [scale=0.45]{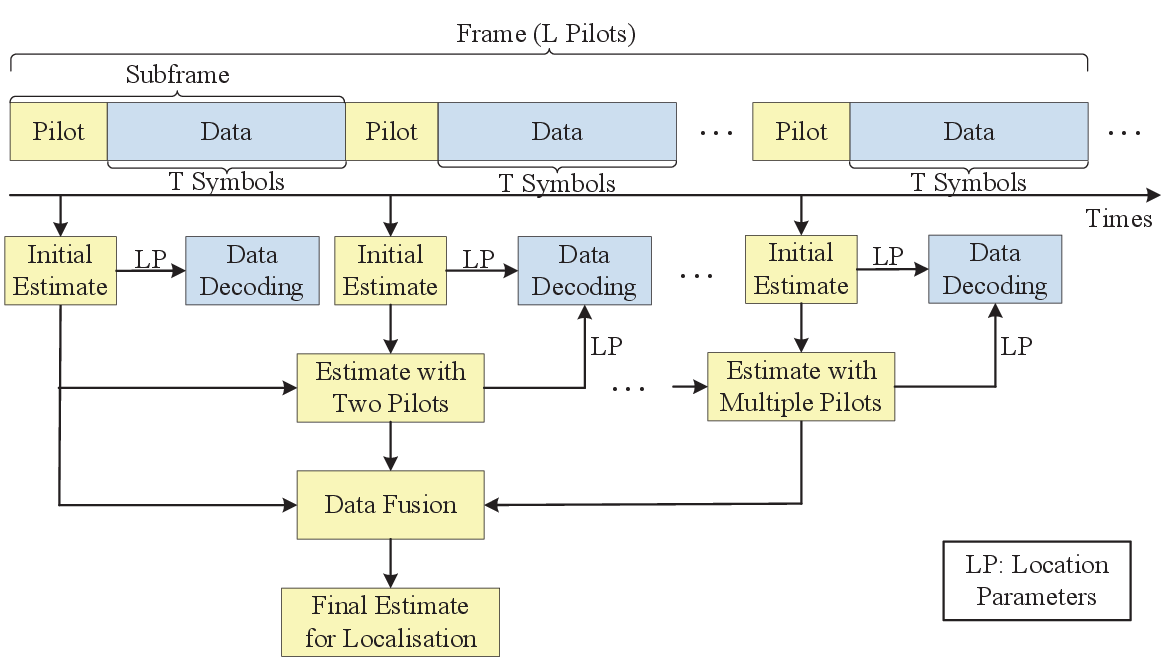}
\caption{The frame and receiver structures for DLDD.}
\label{ISAC diagram}
\end{figure}

\subsection{Motivation and Contributions}

\label{contribution}
It can be seen from the above-surveyed literature that there is a persistent need for developing a more efficient system model compared to traditional ISAC models based on echo signals for localisation, and investigating its functionality under array model imperfections. Motivated by this fact, we introduce a PASCAL system model and evaluate its functionality under model imperfections, as well as we develop simple and effective algorithms for joint localisation and data detection. Unlike almost all introduced ISAC models including \cite{ISACzhen}-\cite{analysis_li}, where the BS emits a dedicated signal towards the targets and receives echos for localisation, the introduced PASCAL system model exploits the fact that drones in general have typical transceiver that can be utilised to send signals actively to a BS for both localisation and communication. As a consequence, this setup is more practical and less energy-consuming than the typical radar concept. To the best of our knowledge, no previous work has investigated the PASCAL system model. The main contributions of this paper can be summarized as follows.
\begin{enumerate}
\item A practical uplink PASCAL system model is introduced and then investigated under a generalised model of gain-phase imperfections. Two main scenarios for PASCAL are considered. The first one is called disjoint localisation and data detection (DLDD), where localisation and data detection are performed sequentially, while in the second one, joint localisation and data detection (JLDD) is pursued.
\item The performance of DLDD is evaluated with gain-phase imperfections, where the average Cramér-Rao lower bound (CRLB) is derived to evaluate the performance limits of the localisation part, while sum data rate (SDR) is employed to assess the communication part. As the exact solution for the average SDR is not tractable, a more accurate approximation method compared to the commonly used first-order Taylor approximation method is proposed to obtain an accurate approximation.
\item For JLDD, the alternating MLE-MLE algorithm and joint MUSIC-MLE algorithm are proposed. Compared to DLDD, the JLDD algorithms manage to achieve improved localisation and
decoding performance by leveraging the location information embedded in the data symbols in addition to the pilot signals. In addition, the proposed algorithm can use more location information while decoding more symbols until all the symbols have been decoded. 
\end{enumerate}
The remainder of this paper is organised as follows. In Sec. II, the system model for PASCAL is presented. Sec. III shows the localisation algorithms and uplink communication with MRC for DLDD. In Sec. IV and Sec. V, the proposed MLE-MLE and MUSIC-MLE algorithms for JLDD and the performance analysis for DLDD are given. Sec. VI provides the numerical results and Sec. VII concludes the paper.

$Notations$: ${[ \cdot ]^T}$,${[ \cdot ]^H}$ and ${[ \cdot ]^*}$ denote the transposition, Hermitian transposition and complex conjugate. $\left \| \cdot \right \|_2$ refers to the Euclidean norm. ${\mathop{\rm var}}( \cdot )$, ${\mathop{\rm cov}}( \cdot )$, $\mathrm{tr}( \cdot )$ and $\Gamma ( \cdot )$ represent the variance, covariance, trace and Gamma function. $\Re ( \cdot )$ indicates the real part of an input argument. $\mathbb{E}[\cdot]$ indicates the statistical expectation. The table of symbols is provided in Table \ref{tableofsymbols}. 

\begin{table*}[t]
{\fontsize{8.8pt}{12pt}\selectfont 
\label{R_1_0}
\caption{ Table of Symbols } 
\centering{}%
\selectfont
\begin{tabular}{|>{\centering}p{0.49cm}|>{\centering}p{4.4cm}|>{\centering}p{0.49cm}|>{\centering}p{4.8cm}|>{\centering}p{0.49cm}|>{\centering}p{4.2cm}|}
\hline 
\textbf{Sym} &
\textbf{Definition} &
\textbf{Sym} &
\textbf{Definition}&\textbf{Sym} &
\textbf{Definition}\tabularnewline
\hline
$K$ & Drones' number & ${{{f}}_{D,k}}$ & Doppler frequency& $v_k$ & Velocity of drone $k$ \tabularnewline
$\phi_k $  & Azimuth angle & $ \theta_k $  & Elevation angle & $\lambda$ & Signal wavelength 
\tabularnewline
$T$ & Symbols in each subframe& ${\bf{  y}}_{t,l}$ &
Received signal vector $t$ in subframe $l$  &  ${\bf{s}}_{t,l}$ &
Symbol vector $t$ in subframe $l$  \tabularnewline
${\bf{A}}$ & Array manifold & ${\boldsymbol{\omega}}$ &
Vector with $f_D$ and path loss  & ${P_k}$ &
Transmitted power of drone $k$ \tabularnewline
${s_{t,l,k}}$ &
\!\!\mbox{Symbol $t$ in subframe $l$ from drone $k$} & ${\bf{s}}_{0,l}$ &
Pilot vector in subframe $l$  & ${\eta _k}$ &
Path loss \tabularnewline
${\alpha }$ & Gain defects & $\Delta {\delta } \vspace{-0.1cm}$ &
Phase defects & $\nu$ &
Location parameter of Rician \tabularnewline
$\sigma_r$ &
Scale parameter of Rician & $\tilde k $ &
Concentration parameter of von Mises & $\tilde \mu $ &
Location parameter of von Mises \tabularnewline
$d_k$ & Distance from BS to drone $k$  & ${\bf{  x}}_{t,l,k}$ &
Processed signal vector with MRC & ${\boldsymbol{\hat {h } }_k}$ &
Estimated channel response
\tabularnewline
${\boldsymbol{ {h } }_k}$ &
Actual channel response of drone $k$  & ${\boldsymbol{ {h } }_p}$ &
Actual channel response of drone $p$ & $R_k$ &
Average data rate of drone $k$ \tabularnewline
${ \gamma _k}$ &
Instantaneous SINR of drone $k$ & $L$ & Subframes'/pilots' number & ${s_{0,l,k}}$ &
Pilot in subframe $l$ from drone $k$ \tabularnewline
\hline 
\end{tabular}
\label{tableofsymbols}
}
\end{table*}

\section{System Model}

This work considers an uplink PASCAL system, where $K$ mobile single-antenna drones, each of which has a Doppler frequency of ${{f_{D,k}}}$ Hz for any $k=\{1,...,K\}$, are deployed in the far field and need to send information signals to a BS  with widely separated antennas. The BS is responsible for localising the drones and decoding their information by using the pilot signals existing in the information packets. Therefore, the considered system is a multiuser single-input multiple-output system (MU-SIMO). It should be noticed that such a model is applicable to Global Positioning System (GPS)-denied environments \cite{GPS1}, for example, suburbs, where precise drone localisation is infeasible. Moreover, the accuracy of commercial GPS might not be guaranteed in drone localisation, especially in estimating the altitude. Thus, drones can be employed to send information signals to the BS actively, while the BS aims to decode uplink data packets in addition to localising these drones. Compared to ground-based vehicles, drones are more suitable for this scenario as drones can easily move close to the BS and fly over the airspace of the BS until finding a proper LOS channel to boost the localisation and communication performance. LoS channel is widely accepted in the literature for UAV-BS links \cite{LOS1} since there is a high probability that strong LOS components for the A2G links exist. Moreover, the LOS channel is adopted in this paper since in radar-based localisation, reflections from the NLOS paths could mislead the localisation process due to the so-called virtual or ghost targets phenomenon \cite{ghost}. Thus, in this paper, algorithms such as ray tracing \cite{ray tracing} are employed prior to localisation to pre-process the received multi-path signal and extract the LoS component that is useful for localisation.

When localising the drones and decoding the transmitted symbols at the BS, similar to \cite{ISACzhen}, we are targeting the scenario in which the velocities and positions (i.e., azimuth angle and elevation angle) of the drones are relatively constant during a frame period. It is worth highlighting that the velocities of drones here are modelled by using the Doppler frequency which is directly related to the drones' velocities through ${{f_{D,k}}}=v_k\cos{\theta_k}/\lambda$, where $v_k$ and $\theta_k $ respectively indicate the velocity and the elevation angle of the $k$th drone, and $\lambda$ denotes the signal wavelength. With this assumption, the azimuth-elevation angles and Doppler frequency are time-invariant during the frame period as their changes can be negligible. The BS employed is composed of a URA with ${M}\times{N}$ antennas, where $M$ and $N$ denote the number of antennas along the $x$-axis and $y$-axis, respectively. Without loss of generality, the distance between any two adjacent antennas of the URA is $d=\lambda /2$. The azimuth angle of the signal received by the BS from the $k$th drone can be denoted by $\phi_k$. Furthermore, we assume that all antennas in the BS suffer from gain-phase errors.

As shown in Fig. \ref{ISAC diagram}, each frame contains $L$ subframes and each subframe contains $T+1$ signals. After receiving the $t$th signal in the $l$th  subframe at the BS and then arranging the outputs of the matched filter, the received signal vector can be written as
\begin{equation} 
\label{received signal matrix2}
 {\bf{  y}}_{t,l} = {\bf{A}}{{\boldsymbol{\omega}}(l)}  {\bf{s}}_{t,l}+ {\bf{  n}}, 
\end{equation}
where ${\bf{  y}}_{t,l} \in \mathbb{C}{^{MN \times 1}}$ contains the signals from all the drones for $t\in\{0,...,T\}$ and $l\in\{1,...,L\}$. ${\bf{A}}{{\boldsymbol{\omega}}(l)}$ denotes the equivalent channel response, in which ${\bf{A}}\in \mathbb{C}{^{MN \times K}}$ indicates the array manifold of the BS, and the vector ${\boldsymbol{\omega}}(l) \triangleq {\mathrm{diag}\{{\eta _1}{{\mathrm{e}}^{j2\pi {f_{D,1}}l/{f_s}}},...,{\eta _K}{{\mathrm{e}}^{j2\pi {f_{D,K}}l/{f_s}}}\}}$ with ${f_s}$ represents the signal sampling frequency and ${\eta _k}$
represents the free space path loss with ${\eta _k}=\frac{\lambda}{4 \pi d_k}$, where $d_k $ refers to the distance from the BS to the $k$th drone. ${\bf{s}}_{t,l}$ refers to the information signal vector with ${\bf{  s}}_{t,l} = [\sqrt{{{P}}_{1}}{s_{t,l,1}},...,\sqrt{{{P}}_{K}}{s_{t,l,K}}]^T$, where ${s_{t,l,k}}$ denotes the $t$th signal within the $l$th subframe from the $k$th drone. In particular, the first signal vector ${\bf{s}}_{0,l}$ refers to the pilot signal vector in the $l$th subframe with ${{{s}}_{0,l,k}}=1$ and thus ${{\bf{s}}_{0,l}}=[ \sqrt{{{P}}_{1}},...,\sqrt{{{P}}_{K}}  ]^T$. ${{{P}}_{k}}$ denotes the transmitted power of the $k$th drone and the power coefficient is defined as ${\varpi _k} = {P_k}/({P_1} + \cdots + {P_K})$. ${{\bf{n}}}$ denotes the additive white Gaussian noise (AWGN) vector. The array manifold ${\bf{A}}$ can be expressed as ${\bf{A}} =\big[{{\bf{\tilde a}}}({\phi _1},{\theta _1}) ,...,{{\bf{ \tilde a}}}({\phi _K},{\theta _K}) ]$, in which the erroneous steering vector of the URA towards the $k$th drone is 
\begin{equation} 
\begin{array}{l}
  {\tilde {\bf{a}}}({\phi _k},{\theta _k}) = [{{{\tilde a}}}_{1,1}({\phi _k},{\theta _k}),...,{{{\tilde a}}}_{M,N}({\phi _k},{\theta _k})]^T,
\end{array}
\end{equation}
where $\displaystyle{{\tilde {a}}}_{m,n}({\phi _k},{\theta _k})\triangleq {\alpha _{m,n}}{e^{j\Delta {\delta _{m,n}}}}{{ {a}}}_{m,n}({\phi _k},{\theta _k})\;\forall \{m,n\}\in \{M,N\}$ with ${\alpha_{m,n} }$ and $\Delta {\delta_{m,n} }$ represent the gain defects and phase defects of the $(m,n)$th antenna, respectively, and ${{ {a}}}_{m,n}({\phi _k},{\theta _k})$ denotes the correct steering vector, which can be written as ${{ {a}}}_{m,n}({\phi _k},{\theta _k})=e^{-j2\pi[({m}-1) d \cos {\phi _k}\sin {\theta _k}+({n}-1) d \sin {\phi _k}\sin {\theta _k}]/\lambda}$. 

\subsection{Gain-Phase Defects Model}

Instead of modelling gain errors and phase errors as mutually independent zero-mean real Gaussian distributions \cite{Performance 3}, \cite{Performance 5}, we consider gain and phase errors as a whole and assume gain-phase defects follow a complex Gaussian distribution.  Thus,  similar to the errors' model in our previous work \cite{aerohan}, we assume that gain imperfections follow a Rician distribution, whereas the phase defects follow a von Mises distribution. It is worth mentioning that the von Mises distribution is 
employed for modelling phase noise because it generalizes the uniform phase distribution, and captures the randomness of actual phase noise in practical conditions \cite{Von}. In line with the assumptions in many references, such as \cite{GPerrors}, \cite{Performance 4}, etc., the gain-phase errors are assumed to be independent across antennas. 

The PDF of the Rician distribution is denoted by
\setcounter{equation}{2}
\begin{equation}
\label{rician PDF}
f({\alpha _{m,n}}) = \frac{{{\alpha _{m,n}}}}{{{\sigma_r ^2}}}{\mathrm{e}}^ {\frac{{ - (\alpha _{m,n}^2 + {\nu ^2})}}{{{2{\sigma_r ^2}}}}}{I_0}\Big(\frac{{{\alpha _{m,n}}\nu }}{{{\sigma_r ^2}}}\Big), 
\end{equation}
where $\nu$ and $\sigma_r$ represent the location and scale parameters of the Rician distribution, respectively and $\displaystyle {I_0}(x)$ is the modified Bessel function of the first kind with zero order. 

As phase errors follow the independent von Mises distribution, the PDF can be denoted by 
\begin{equation}
\label{e60}
 f(\Delta {\delta _{m,n}}) = \frac{1}{({{{2\pi {I_0}(\tilde k)}}})}{\mathrm{e}}^{\tilde k\cos (\Delta {\delta _{m,n}} - \tilde \mu )},
\end{equation}
where $\tilde \mu $ and $\tilde k $ denote the measures of location and concentration. As the ideal case of phase defects is zero (e.g. $\Delta {\delta _{m,n}}=0 $), the mean value of the phase error is typically $\tilde \mu=0 $.  

In the PASCAL system, two main scenarios, namely DLDD and JLDD, are considered in Secs. III and IV, respectively. In DLDD, the entire process at the BS is divided into two stages, where the localisation is performed in the first stage and the communication is conducted in the second stage. On the other hand, in JLDD, localisation and data detection at the BS are conducted jointly and two algorithms including the alternating MLE-MLE and joint MUSIC-MLE algorithms are proposed.

\section{Disjoint localisation and data detection}
\label{locsalition algorithm}
\label{deriv_ap}

In DLDD, as shown in the frame structure in Fig. \ref{ISAC diagram}, each frame is composed of $L$ subframes and each subframe contains one pilot and $T$ symbols, where the pilot signal is employed to provide initial estimates for the location parameters in each subframe, which will be used for decoding data symbols in that subframe. After estimating the parameters by using $L$ pilots, the final estimates for location parameters will be obtained. However, to improve the decoding performance and capacity, the BS can also wait until the final estimation for the location parameters is completed by using all the pilots in the frame and then the transmitted symbols are decoded for all the $L$ subframes. Albeit, this may increase the system's latency, and thus a trade-off should be considered.
\subsection{Localisation Stage}
\label{MLE localisation}
To begin, we aim to derive the MLE for DLDD with gain-phase errors. By using ${L}$ pilots during the whole frame, as shown in Fig. \ref{ISAC diagram}, the received signal vector ${{\bf{\bar y}}}$ is obtained as 
\begin{equation}
\label{X1}
{{\bf{\bar y}}} = {\left\{\{{{\bf{A}}}{\boldsymbol{\omega}}(1){{\bf{s}}_{0,1}}\}^T,...,\{{{\bf{A}}}{\boldsymbol{\omega}}({L}){{\bf{s}}_{0,L}}\}^T\right\}^T} + {{\bf{\bar n}}}, 
\end{equation}
where ${{\bf{\bar y}}}\in {{\mathbb{C}}^{{M}{N}{L} \times 1}}$. The variables in ${{\bf{\bar y}}}$ include the gain-phase defects in ${{\bf{A}}}$ and AWGN noise. As they follow a complex Gaussian distribution, ${{\bf{\bar y}}}$ follows a multivariate Gaussian distribution and its PDF conditioned on $\boldsymbol{\beta}$ can be denoted by
\begin{equation}
\label{tilde y PDF}
f({{\bf{\bar y}}}|\boldsymbol{\beta} ) = {1}/\left({{{\pi ^{MN{L}}}\det ({\bf{\Gamma }})}}\right){{e}^{ - {{\left[{{\bf{\bar y}}} - \boldsymbol{\mu} \right]}^H}{{\bf{\Gamma }}^{ - 1}}\left[{{\bf{\bar y}}} - \boldsymbol{\mu} \right]}},
\end{equation}
where $\boldsymbol{\beta} = {[{{\boldsymbol{\phi }}^T},{{\boldsymbol{\theta }}^T},{{\boldsymbol{f_{D}}}^T}]^T}$, which refers to the vector consisting of the deterministic unknown location parameters with $\boldsymbol \phi=[\phi_1,...,\phi_K]$, $\boldsymbol \theta=[\theta_1,...,\theta_K] $ and $ \boldsymbol{f_{D}} =[f_{D,1},...,f_{D,K}] $.  

\begin{figure}[ptb]
\centering 

{\includegraphics  [height=1.5in, width=2.736in]{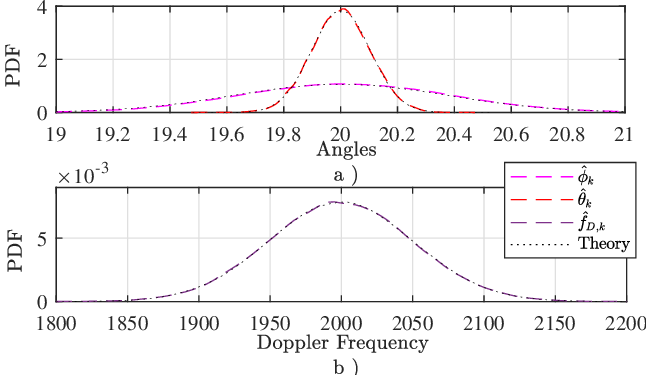}}

 \caption{PDF of estimated a) angles and b) Doppler frequency under gain-phase defects.} 
\label{PDF}
\end{figure}

The MLE, which can identify the parameter values that optimise the log-likelihood function over the parameter space, is
\begin{equation}
\label{True ML}
{\left[ {\boldsymbol{ \phi} , \boldsymbol{\theta} ,\boldsymbol{f_{D}} } \right] = \mathop {\arg \max }\limits_{\boldsymbol{ \phi} , \boldsymbol{\theta} ,\boldsymbol{f_{D}} } \ln f({\bf{\bar y}}|{\boldsymbol{\beta }})},
\end{equation}

 \begin{figure*}[!b]
\hrulefill
\setcounter{equation}{10}
 \begin{align}
\label{gammaij}
 \!\!\!\!\!\!\!\!  {\boldsymbol{\Gamma }}_{i,j}\!\!&=\!\!{\mathbb{E}}[{\alpha _{m_1,n_1}}\!{\alpha _{m_2,n_2}}]{\mathbb{E}}[{e^{j(\Delta {\delta _{m_1,n_1}} \!-\! \Delta {\delta _{m_2,n_2}})}}]{Q_{ {m_1}{n_1}{l_1}}}\!Q_{ {m_2}{n_2}{l_2}}^ * \!\!\!- \!\!{\mathbb{E}}[{\alpha _{m_1,n_1}}]{\mathbb{E}}[{e^{j\Delta {\delta _{m_1,n_1}}}}]{{{\mu}}^*_{{m_2}{n_2}{l_2}}}\!{Q_{ {m_1}{n_1}{l_1}}}\!\!\!-\!\!{\mathbb{E}}[{{\bar n}_{{m_1}{n_1}{l_1}}}\!{{\bar n}_{{m_2}{n_2}{l_2}}}],\!\!\!\!\!
\end{align} 
\end{figure*}

By removing constant terms which do not impact the maximization operation,\eqref{True ML} can be simplified to
\setcounter{equation}{7}
\begin{equation}
\label{True ML2}
\begin{array}{*{20}{l}}
{\left[ {\boldsymbol{ \phi} , \boldsymbol{\theta} ,\boldsymbol{f_{D}} } \right] = \mathop {\arg \min }\limits_{\boldsymbol{ \phi} , \boldsymbol{\theta} ,\boldsymbol{f_{D}} } {({\bf{\bar y}} - \boldsymbol{\mu} )^H}\bf{\Gamma} ^{ - 1}({\bf{\bar y}} - \boldsymbol{\mu} )  }\\ \ \ \ \ \ \ \ \ \ \ \ \ \
= \mathop {\arg \min }\limits_{\phi ,{\boldsymbol{\theta }},\boldsymbol{f_{D}}} ||{\bf{\bar y}} - {\boldsymbol{\mu }}||_2^2,
\end{array}
\end{equation}
where the covariance matrix ${\bf{\Gamma }}$ and mean vector $\boldsymbol{\mu}$ can be respectively calculated by ${\bf{\Gamma }}={\mathbb{E}}[({{\bf{\bar y}}}-{\boldsymbol{\mu}})({{\bf{\bar y}}}-{\boldsymbol{\mu}})^H]$ and 
\begin{equation}
\label{mu}
\begin{array}{l}
\!\!\!\! \boldsymbol{\mu}=
{\left\{\{{{\mathbb{E}}[{\bf{A}}}{\boldsymbol{\omega}}(1){{\bf{s}}_{0,1}}]\}^T,...,\{{\mathbb{E}}[{{\bf{A}}}{\boldsymbol{\omega}}({L}){{\bf{s}}_{0,L}}]\}^T\right\}^T},
\end{array}
\end{equation}
where $\boldsymbol{\mu}\in {{\mathbb{C}}^{{M}{N}{L} \times 1}}$ and the general expression for the $mn{l}$th element of $\boldsymbol{\mu}$ can be given by ${{\mu}}_{mn{l}} = Q_{mnl}{\mathbb{E}}[{\alpha _{m,n}}]{\mathbb{E}}[{e^{j\Delta {\delta _{m,n}}}}] $, in which the closed-form expressions of $\mathbb{E}[{\alpha _{m,n}}]$ and $\mathbb{E}[{e^{j\Delta {\delta _{m,n}}}}]$ can be obtained by using Theorem 1 on page \pageref{theorem 1} and the full derivation in Appendix A, respectively, and $Q_{mnl}$ can be written as 
\begin{align}
   Q_{mnl}=\sum\limits_{k = 1}^K \eta_k \sqrt{{{P}}_{k}} {{{a}}}_{m,n}({\phi _k},{\theta _k})e^{j2\pi {f_{D,k}}l/{f_s} }. 
\end{align}

The covariance matrix ${\bf{\Gamma }}\in {{\mathbb{C}}^{{M}{N}{L} \times {{M}{N}{L}}}}$ in  \eqref{True ML2} is also necessary for the derivation of CRLB, where the $(i,j)$th element in ${\boldsymbol{\Gamma }}$ for $i,j=\{1,...,MN{L}\}$ can be denoted by $ {\boldsymbol{\Gamma }}_{i,j}={\mathbb{E}}[({{\bar y}}_{m_1n_1{l_1}}-{{\mu}}_{m_1n_1{l_1}})({{\bar y}}_{m_2n_2{l_2}}-{{\mu}}_{m_2n_2{l_2}})^*]$, where ${{\bar y}}_{m_1n_1{l_1}}$ and ${{\bar y}}_{m_2n_2{l_2}}$ can be expressed by using the general expression ${{\bar y}}_{mn{l}} = Q_{mnl} {\alpha _{m,n}}{e^{j\Delta {\delta _{m,n}}}}+\bar n_{mnl}  $, while ${{\mu}}_{m_1n_1{l_1}}$ and ${{\mu}}_{m_2n_2{l_2}}$ can also be  denoted by using their general expression ${{\mu}}_{mn{l}}$. By expanding the product inside the expected value in ${\boldsymbol{\Gamma }}_{i,j}$ and discarding the terms containing ${\mathbb{E}}[{\bar n}_{mnl}]$ as ${\mathbb{E}}[{\bar n}_{mnl}]=0$, the simplified result of ${\boldsymbol{\Gamma }}_{i,j}$ can be obtained, which is shown in \eqref{gammaij} on top of page \pageref{gammaij}. 

To derive ${\mathbb{E}}[{\alpha _{m_1,n_1}}{\alpha _{m_2,n_2}}]$ and ${\mathbb{E}}[{e^{j(\Delta {\delta _{m_1,n_1}} - \Delta {\delta _{m_2,n_2}})}}]$ in \eqref{gammaij}, the independence of gain-phase defects across different antennas is employed and two cases need to be considered. When $m_1 \ne m_2 $ or $n_1 \ne n_2 $,
 \setcounter{equation}{11}
\begin{subequations}
 \begin{equation}
    \ \ \ \ \ {\mathbb{E}}[{\alpha_{m_1,n_1}}{\alpha_{{m_2},{n_2}}}]={\mathbb{E}}[{\alpha_{m_1,n_1}}]{\mathbb{E}}[{\alpha_{{m_2},{n_2}}}],   
    \vspace{0.05cm}
 \end{equation} 
 \begin{equation}
    {\mathbb{E}}[{e^{j(\Delta {\delta _{m_1,n_1}}-\Delta {\delta _{{m_2},{n_2}})}}}]={\mathbb{E}}[{e^{j\Delta {\delta _{m_1,n_1}}}}]{\mathbb{E}}[{e^{-j\Delta {\delta _{{m_2},{n_2}}}}}],        
 \end{equation}
\end{subequations}  
where ${\mathbb{E}}[{\alpha_{m_1,n_1}}]$ and  ${\mathbb{E}}[{\alpha_{{m_2},{n_2}}}]$ can be derived by using Theorem 1 on page \pageref{theorem 1}. Thereafter, the closed-form expressions of $ {\mathbb{E}}[{e^{j\Delta {\delta _{m_1,n_1}}}}]$ and $ {\mathbb{E}}[{e^{-j\Delta {\delta _{{m_2},{n_2}}}}}]$ can be calculated by utilising the derivation of their general expression ${\mathbb{E}}\big[{e^{\pm jC\Delta {\delta _{\rho}}}}\big]$ in Appendix A. On the other hand, when $m_1  = m_2 $ or $n_1 = n_2 $, ${\mathbb{E}}[{\alpha_{m_1,n_1}}{\alpha_{{m_2},{n_2}}}]={\mathbb{E}}[{\alpha_{m,n}^2}] $ and ${\mathbb{E}}[{e^{j(\Delta {\delta _{m_1,n_1}}-\Delta {\delta _{{m_2}, { n_2}})}}}]=1$, where ${\mathbb{E}}[{\alpha_{m,n}^2}]$ can be obtained by Theorem 1.

By substituting the calculated results of ${\mathbb{E}}[{\alpha _{m_1,n_1}}{\alpha _{m_2,n_2}}]$ and ${\mathbb{E}}[{e^{j(\Delta {\delta _{m_1,n_1}} - \Delta {\delta _{m_2,n_2}})}}]$ into \eqref{gammaij} and also employing the closed-form expressions of $\mathbb{E}[{\alpha _{m,n}}]$ and $\mathbb{E}[{e^{j\Delta {\delta _{m,n}}}}]$, the mean vector $\boldsymbol{\mu}$ and covariance matrix ${\bf{\Gamma }}$ can be obtained, which are then substituted into \eqref{True ML2} to provide the estimation of azimuth angle, elevation angle and Doppler frequency for MLE when gain-phase errors are considered.

\textbf{\textit{Theorem 1}}:
\label{theorem 1}
The $c$-th central moment of the Rician distribution is given by
\begin{equation}
\begin{array}{l}
{\mathbb E}[\alpha _{{\rho}}^c] \!\!=\!\!\displaystyle \frac{{\mathrm{e}}^ {\!\! - \frac{{{\!\nu ^2}}}{{2{\sigma_r ^2}}}}}{{2{\sigma_r ^2}}} \!\!\sum\limits_{b = 0}^\infty \!\! {\frac{{\nu^{2b}}{\!(\!\frac{1}{{2{\sigma_r ^2}}}\!)^{b - \frac{c+2}{2}}}}{{b!\Gamma (b\! +\! 1)}}}  \Gamma \left(\!\frac{2b\!+\!c\!+\!2}{2} \right)\!\!,\!\!\!\!
\end{array} \label{E-theorem-1}
\end{equation}
where $\gamma(\cdot)$ denotes a lower incomplete gamma function and $\rho\in \{\{m ,n\}, \{m_1 ,n_1\}, \{{m_2},{n_2}\} ,\{{m_3},{n_3}\}, \{{m_4},{n_4}\}\}$.
\textit{Proof: See Appendix A.}

\subsection{ Communication Stage}
\label{com stage}
In DLDD, the estimated location parameters by MLE in Sec. \ref{locsalition algorithm}-\ref{MLE localisation} are utilised to infer the channel responses in the communication stage. Afterwards, the MRC technique is adopted to combine the signals using the estimated channel responses (i.e., ${\bf{\hat H}}={\bf{ {\hat A } }}{\boldsymbol{\hat \omega}}(l)$) to improve the received signal quality. The communication signal after being processed using the MRC technique shown in \cite{approx1}, i.e., ${\bf{ x}}_{t,l}\in \mathbb{C}{^{K \times 1}}$, can be written as 
\begin{equation}
\label{x0_0}
{\bf{ x}}_{t,l}  
 =[{\bf{ {\hat A } }}{\boldsymbol{\hat \omega}}(l)]^H {\bf{ A }}{\boldsymbol{ \omega}}(l)  {\bf{s}}_{t,l}+ [{\bf{ {\hat A } }}{\boldsymbol{\hat \omega}}(l)]^H {\bf{  n}} ,
\end{equation}
where ${\bf{ \hat A}} = \big[{{\bf{ a}}}({\hat \phi _1},{\hat \theta _1}) ,...,{{\bf{  a}}}({\hat \phi _K},{\hat \theta _K}) ]$ and ${\boldsymbol{ \hat \omega}}(l) \triangleq \mathrm{diag}({{\mathrm{e}}^{j2\pi {\hat f_{D,1}}l/{f_s}}},...,{{\mathrm{e}}^{j2\pi {\hat f_{D,K}}l/{f_s}}} )$. The estimated steering vector associated with the $k$th drone can be given by
\begin{equation}
{{\bf{ a}}}({\hat \phi _k},{\hat \theta _k})\!\!=\!\!
\{1,...,{e^{- j2\pi  [({M}-1)d\cos {\hat \phi _k}+({N}-1)d\sin {\hat \phi _k}]\sin {\hat \theta _k}/\lambda}}\} ,
\end{equation}
where ${\hat \phi _k}$, ${\hat \theta _k} $ and ${{\hat f}_{D,k}}$ indicate the azimuth-elevation angles and Doppler frequency estimated by the localisation algorithms. 

  \begin{figure}[ptb]
\centering
\includegraphics [scale=0.46]{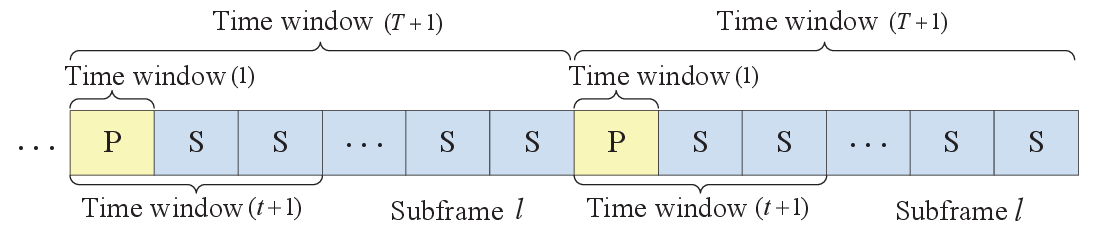}
\caption{ The diagram for JLDD algorithms }
\label{MUSIC-MLE}
\end{figure} 

Interestingly, we find that the PDF of the estimated azimuth angle, elevation angle and Doppler frequency follow real Gaussian distribution under gain-phase defects. Take MLE as an example, by collecting samples of the estimated location parameters in $10^6$ simulation tests, their PDFs are plotted in Fig. \ref{PDF}. In Fig. \ref{PDF}, we consider a drone, which is located at $({\phi_1},{\theta_1},{f_{D,1}}) = [({20^ \circ },{20^ \circ },2000\; \mathrm{Hz})$, at $\mathrm{SNR}= 8\;\mathrm{dB}$. The simulation parameters are summarized in Table \ref{sim par}. The location and scale parameters of the Rician distributed gain defects are set to $\nu=0.5$ and $\sigma_r=1$, and the concentration parameter of the von Mises distributed phase defects is $\tilde k=1000$. The considered BS is composed of $M\times N=6\times 6$ antennas and the number of pilots is $L=10$ in the whole frame, where each pilot signal is inserted in each subframe of $T=100$ symbols.  Thereafter, the mean and variance of the estimated parameters are calculated and the PDF of the estimated location parameters is compared with the theoretical Gaussian PDF with the same mean and variance as those of the estimated parameters. As shown in Fig. \ref{PDF}, the PDFs of the estimated location parameters and the theoretical Gaussian distribution match very well, which indicates these estimated parameters follow the Gaussian distributions under gain-phase defects. Moreover, the PDFs of the estimated parameters have means that are very close to the actual values of location parameters, which implies that the derived MLE is unbiased. 

In \eqref{x0_0}, the $k$th element of ${\bf{ x}}$ can be written as
\begin{equation}
       {{ x}}_{t,l,k}\!\! = \!\!\sqrt{{{P}}_{K}}{{\boldsymbol{\hat {h } }^H_k}} {{\boldsymbol{ {h } }_k}} {{s_{t,l,k}}}\!+\!\!\!\!\!\!\sum\limits_{ p = 1,p \ne k}^K \!\!\!\!\!\!\sqrt{{{P}}_{p}}{{\boldsymbol{\hat {h } }^H_k}} {{\boldsymbol{ {h } }_p}} {{s_{t,l,p}}}\!+ \!{{\boldsymbol{\hat {h } }^H_k}}{\bf{ n}}, 
\end{equation}
where ${\boldsymbol{\hat {h } }_k}$ denotes the erroneous estimated channel response of the $k$th drone, while ${{\boldsymbol{ {h } }_k}} $ and ${{\boldsymbol{ {h } }_p}} $ indicate the actual channel response corresponding to the $k$th drone and the $p$th drone. ${\boldsymbol{\hat {h } }_k}$ can be represented by $ {\boldsymbol{\hat {h } }_k}= {{\bf{   a}}}({\hat \phi _k},{\hat \theta _k}) {{\mathrm{e}}^{j2\pi {\hat f_{D,k}}l/{f_s}}}$, while ${{\boldsymbol{ {h } }_k}} $ and ${{\boldsymbol{ {h } }_p}} $ can be denoted by the general expression as $\displaystyle {\boldsymbol{ {h } }_\varsigma}= \eta_k{{\bf{ \tilde a}}}({ \phi _\varsigma},{ \theta _\varsigma}) {{\mathrm{e}}^{j2\pi { f_{D,\varsigma}}l/{f_s}}}$, where $\varsigma \in\{k,p\}$. 

Afterwards, the performance analysis for both localisation and communication of DLDD is provided in Sec. \ref{performance analysis DLDD}.

\section{Joint localisation and data detection}
In JLDD, instead of performing communication after localisation like in DLDD, the localisation and data detection are conducted jointly, and the location information in the data symbols is also employed instead of using pilots only. For this, two novel algorithms referred to the alternating MLE-MLE and joint MUSIC-MLE algorithms are proposed. As shown in Fig. \ref{MUSIC-MLE}, the estimation of location parameters and symbols in each subframe are carried out in the time windows, where time window ($t+1$) for $t \leq T$ indicates the time window of a length of $t+1$ as it contains one pilot signal and $t$ data symbols. Note that even though Fig. \ref{MUSIC-MLE} is an example of a single-drone case, the proposed algorithms here are suitable for multiple-drone scenarios. As the length of the time window increases, the localisation performance is enhanced since the algorithms can use more location information from more symbols as the data symbols themselves in our system model also contain location information. Therefore, by enhancing the localisation performance, the decoding performance is improved as the localisation performance can directly affect the decoding performance.  This proposed concept is similar to increasing the number of pilots to enhance the system performance, however, the proposed algorithms utilise one pilot signal only, which makes the algorithms very effective for integrated communication and localisation systems.

\begin{algorithm}[tp] 
	\caption{The alternating MLE-MLE algorithm}
	\label{Al1}
    \vspace{0.1cm}
  \textbf{\mbox{Time window ($1$) (Pilot signal):}}
   \vspace{-0.4cm}
	\begin{algorithmic}[1]    
	\STATE {\bf Initialization:} Set $\boldsymbol{\hat \phi}^{(0)}$, $\boldsymbol{\hat \theta}^{(0)}$,${\boldsymbol f}^{(0)}_{\boldsymbol{D}}$ \\
       \STATE Calculate $\boldsymbol{\hat \phi}^{(z)}$ and $\boldsymbol{\hat \theta}^{(z)}$ based on  ${\boldsymbol{ \hat f}}^{(z-1)}_{\boldsymbol{D}}$ in \eqref{phi_theta1}.
	\STATE Calculate  ${\boldsymbol{\hat f} }^{(z)}_{\boldsymbol{D}}$ based on $\boldsymbol{\hat \phi}^{(z-1)}$ and $\boldsymbol{\hat \theta}^{(z-1)}$ in \eqref{s_f1}. 
	     \STATE until the above alternative optimisation convergences,
       \STATE {{\bf output:}} $\boldsymbol{\hat \phi_0}$, $\boldsymbol{\hat \theta_0}$, ${\boldsymbol{ \hat f}}_{\boldsymbol{D,0}}$.
          	\end{algorithmic}
 \textbf{\mbox{ Time window ($t+1$) (Pilot signal and $t$ symbols):}}
 \vspace{-0.4cm}
	\begin{algorithmic}[1]    
	\STATE {\bf Initialization:} Set $\boldsymbol{\hat \phi}^{(0)}$, $\boldsymbol{\hat \theta}^{(0)}$,${\boldsymbol f}^{(0)}_{\boldsymbol{D}}$ and ${\bf{  \hat s}}^{(0)}_{t,l}$\\
       \STATE Calculate $\boldsymbol{\hat \phi}^{(z)}$, $\boldsymbol{\hat \theta}^{(z)}$ based on ${\bf{  \hat s}}^{(z-1)}_{t,l}$ and ${\boldsymbol{ \hat f}}^{(z-1)}_{\boldsymbol{D}}$ in \eqref{phi_theta1}.
	\STATE Calculate ${\bf{ \hat s}}^{(z)}_{t,l}$, ${\boldsymbol{\hat f} }^{(z)}_{\boldsymbol{D}}$ based on $\boldsymbol{\hat \phi}^{(z-1)}$ and $\boldsymbol{\hat \theta}^{(z-1)}$ in \eqref{s_f1}. 
	     \STATE until the above alternative optimisation convergences,
       \STATE {{\bf output:}} $\boldsymbol{\hat \phi_{t}}$, $\boldsymbol{\hat \theta_{t}}$, ${\boldsymbol{ \hat f}}_{\boldsymbol{D,{t}}}$ and ${\bf{  \hat s}}_{\boldsymbol{t,l}}$.
          	\end{algorithmic}
 \textbf{\mbox{ Time window ($T+1$) (Pilot signal and $T$ symbols):}}
       \begin{algorithmic}[1]
       \STATE Obtain the final estimation of location parameters and detect the last symbol using the same process as that of time window ($t+1$),
       \STATE {{\bf output:}} $\boldsymbol{\hat \phi_{T}}$, $\boldsymbol{\hat \theta_{T}}$, ${\boldsymbol{ \hat f}}_{\boldsymbol{D,{T}}}$ and ${\bf{  \hat s}}_{\boldsymbol{T,l}}$.
       \end{algorithmic}
\end{algorithm}

\subsection{Alternating MLE-MLE Algorithm}
In the alternating MLE-MLE algorithm, the estimation for azimuth-elevation angles and the estimation for the $t$th symbol and Doppler frequency are performed alternatively
using two MLEs until the convergence is achieved at the time window with a length of $t+1$. Define a new signal vector ${\bf{ y}}_1=[{\bf{  y}}_{0,l}^T,..., {\bf{ y}}_{t,l}^T]^T $, which contains $t+1$ signals in the $l$th subframe and ${\bf{  y}}_{0,l}$ contains the pilot signal. The first phase of the alternating MLE-MLE algorithm is to estimate azimuth angle and elevation angle given ${\boldsymbol {\hat f}}^{(z-1)}_{\boldsymbol{D}}$ and ${\bf{\hat s}}^{(z-1)}_{t,l} $, where $z$ denotes the iteration index of the alternative optimisation, as
\begin{equation}
\label{phi_theta1}
\!\!\!\!\!\!  \left[\boldsymbol{\hat  \phi}^{(z)}, \boldsymbol{\hat  \theta}^{(z)} \right] \!\! =\!\! \mathop {\arg \min }\limits_{\boldsymbol{\phi}^{(z)}, \boldsymbol{\theta}^{(z)}} \left|\left|{{\bf{ y}}_1}\!\! -\!\!
\begin{bmatrix} \!\!\!\!
{\mathbb{E}}[{{\bf{A}}}{\boldsymbol{\omega}}({\boldsymbol {\hat f}}^{(z-1)}_{\boldsymbol{D}},l){{\bf{ s}}_{0,l}}]\\
 \!\!\!\!{\mathbb{E}}[{{\bf{A}}}{\boldsymbol{\omega}}({\boldsymbol {\hat f}}^{(z-1)}_{\boldsymbol{D}},l){{\bf{\hat s}}_{1,l}}]\\
\vdots
\\ \ \ \  
 \!\!\!\! {\mathbb{E}}[{\bf{A}}{\boldsymbol{\omega}}({\boldsymbol {\hat f}}^{(z-1)}_{\boldsymbol{D}},l)  {\bf{\hat s}}^{(z-1)}_{t,l}]
\end{bmatrix} \right|\right|^2_2, 
\end{equation}
where $ \boldsymbol{\hat f_{D}} \!\!=\!\![\hat f_{D,1},...,\hat f_{D,K}] $, ${{\bf{ s}}_{0,l}}\!\!=\!\!1$ and ${{\bf{\hat 
 s}}_{1,l}},...,{{\bf{\hat 
 s}}_{t-1,l}}$ have been estimated in previous time windows when the convergence has been reached, thus they are known in the current time window. The expectation calculation is the same as the method used in MLE in Sec. \ref{locsalition algorithm}-\ref{MLE localisation}. It is noteworthy observing that our proposed algorithm is universal regardless of the presence of gain-phase defects since these expected values become constant values in the absence of gain-phase defects.

Afterwards, the second phase of alternating MLE-MLE is to estimate Doppler frequency and ${\bf{ \hat s}}_{t,l}$ given ${\boldsymbol{\hat \phi}}^{(z-1)}$ and ${\boldsymbol{\hat \theta}}^{(z-1)} $ as 
\begin{equation} 
\label{s_f1}
\!\!\!\!\!\!\!\! \left[{\bf{ \hat s}}^{(z)}_{t,l}\!, \!{\boldsymbol {\hat f}}^{(z)}_{\boldsymbol{D}} \right] \!\! =\!\! \mathop {\arg \min }\limits_{{\bf{  s}}^{(z)}_{t,l}, {\boldsymbol f}^{(z)}_{\boldsymbol{D}}} \left|\left|{{\bf{ y}}_1}\!\! -\!\!
\!\! \begin{bmatrix} 
{\mathbb{E}}[{{\bf{A}}}({\boldsymbol{\hat \phi}}^{(z\!-\!1)}\!,\! {\boldsymbol{\hat \theta}}^{(z-1)}){\boldsymbol{\omega}}(k){{\bf{ s}}_{0,l}}]\\
{\mathbb{E}}[{{\bf{A}}}({\boldsymbol{\hat \phi}}^{(z\!-\!1)}\!,\! {\boldsymbol{\hat \theta}}^{(z-1)}){\boldsymbol{\omega}}(k){{\bf{\hat s}}_{1,l}}]\\
\vdots\\ \!\! \! 
{\mathbb{E}}[{\bf{A}}({\boldsymbol{\hat \phi}}^{(z\!-\!1)}\!,\!{\boldsymbol{\hat \theta}}^{(z\!-\!1)}){\boldsymbol{\omega}}(l)  {\bf{s}}_{t,l}]
\end{bmatrix}\!\!\right|\right|^2_2\!\!,
\end{equation}
where ${\boldsymbol {\hat \phi}}=[{\hat \phi}_1,...,{\hat \phi}_K]$ and ${\boldsymbol {\hat \theta}}=[{\hat \theta}_1,...,{\hat \theta}_K] $. It should be noticed that \eqref{phi_theta1} and \eqref{s_f1} are solved using exhaustive search and one symbol is detected in each time window.
 
The outputs of \eqref{phi_theta1} and \eqref{s_f1} when the convergence is reached by satisfying the criterion as $||{\boldsymbol{\hat \beta}}^{(z)}-{\boldsymbol{\hat \beta} }^{(z-1)} ||_2< \epsilon$, where ${\boldsymbol{\hat \beta}}^{(z)} \in \{{\boldsymbol{\hat \phi}}^{(z)},{\boldsymbol{\hat \theta}}^{(z)},{\boldsymbol {\hat f}}^{(z)}_{\boldsymbol{D}}\}$, can provide the estimation for both the location parameters and symbol ${\bf{ \hat s}}_{t,l}$ in the time window with a length of $t+1$. In the time window with a length of $T+1$, the alternating MLE-MLE algorithm can provide the final estimation for the location parameters as the location information from all symbols in the current subframe has been used up, whereas the decoding is conducted symbol by symbol to minimise the latency. It is also worth mentioning that the signals in the subframe contain the same $\boldsymbol{\phi}$, $\boldsymbol{\theta}$ and ${\boldsymbol {f}}_{\boldsymbol{D}}$ as illustrated in the system model in Sec. II. However, the symbols in each signal are different, which leads to different and independent received signals which creates some sort of diversity when estimating the location parameters. In addition, as shown in \eqref{s_f1}, the estimation accuracy of $\boldsymbol{\phi}$ and $\boldsymbol{\theta}$ will affect the estimation of the symbol, and thus a better localisation performance leads to a more accurate data detection. A detailed description of alternating MLE-MLE can be found in {Algorithm 1}, where $\boldsymbol{\hat \phi_{t}}$, $\boldsymbol{\hat \theta_{t}}$ and ${\boldsymbol{ \hat f}}_{\boldsymbol{D,{t}}}$ indicate the estimated location parameters in the time window with a length of $t+1$.

 \begin{figure*}[!b]
\hrulefill
\label{gammax}
\setcounter{equation}{35}
\begin{align}
\label{gammax}
\!\!\!\!\!\!\!\!\!  {\mathbb{E}}[|{{\boldsymbol{\hat {h } }^H_{k}}} \!{{\boldsymbol{ {h } }_{\varsigma}}}{|^2}]  
\!\!=\!\! \!\!\!\!\!\!\!\!{\sum\limits_{m_1\!,\!m_2 \in S_1} }\!{\sum\limits_{n_1\!,\!n_2 \in S_2} } \!\!\!\!\eta^2_\varsigma {\mathbb{E}}[{\alpha_{m_1,n_1}}\!{\alpha_{{ m_2},{ n_2}}}]   {\mathbb{E}}[{e^{j(\Delta {\delta _{m_1,n_1}}\!-\!\Delta {\delta _{{ m_2},{ n_2}})}}}]{{{   a}_{m_1,n_1}}}\!({ \phi _\varsigma}\!,\!{ \theta _\varsigma}){{{   a}^*_{m_2,n_2}}}\!({ \phi _\varsigma}\!,\!{ \theta _\varsigma})\!\underbrace{{\mathbb{E}}[{{{   a}^*_{m_1,n_1}}}\!({ \hat \phi _k}\!,\!{\hat \theta _k}){{{   a}_{m_2,n_2}}}\!({\hat \phi _k}\!,\!{\hat \theta _k})]}_{{\mathbb{E}}_1},\!\!\!\!\!
\end{align} 
\end{figure*}

\subsection{Joint MUSIC-MLE Algorithm}

On the other hand, in the joint MUSIC-MLE algorithm, the MUSIC algorithm is applied first to estimate the azimuth-elevation angles and then MLE is adopted to detect the symbol and estimate the Doppler frequency. In the time window with a length of $t+1$, the matrix form of ${{\bf{ y}}_1}$ is constructed first as ${\bf{ Y}}_1=[{\bf{  y}}_{0,l},..., {\bf{ y}}_{t,l}]={{\bf{A}}(\boldsymbol{\phi},\boldsymbol{\theta})}{\bf{W}} $, where ${\bf{W}}=[{\boldsymbol{\omega}}({\boldsymbol { f}}_{\boldsymbol{D}},l){{\bf{ s}}_{0,l}},...,{\boldsymbol{\omega}}({\boldsymbol { f}}_{\boldsymbol{D}},l){\bf{ s}}_{t,l}] $. To apply MUSIC, the covariance matrix ${\bf{ R}}$ of ${\bf{ Y}}_1$ needs to be calculated as ${\bf{ R}}={{\bf{A}}(\boldsymbol{\phi},\boldsymbol{\theta})}{\bf{R}_W}{{\bf{A}}^H(\boldsymbol{\phi},\boldsymbol{\theta})}+\sigma^2{\bf{I}}$, where ${\bf{R}_W}\triangleq \mathbb{E}[{\bf{ W}}{{\bf{ W}} ^H}]$ and then the noise subspace can be obtained by using the eigendecomposition method for the covariance matrix ${\bf{ R}}$ as
\setcounter{equation}{18}
\begin{equation}
   {\bf{ R}} = {{\bf{U}}_s}{{\bf\Sigma} _s}{\bf{U}}_s^H + {{\bf{U}}_n}{{\bf\Sigma} _n}{\bf{U}}_n^H, 
\end{equation}
where ${{\bf{U}}_s} $ and ${{\bf{U}}_n} $ denote the signal subspace and noise subspace, and ${{\bf\Sigma} _s}$ and ${{\bf\Sigma} _n}$ represent the diagonal matrices. By observing that ${{\bf{A}}}(\boldsymbol{\phi},\boldsymbol{\theta}){\bf{R}_W}{{\bf{A}}^H(\boldsymbol{\phi},\boldsymbol{\theta})}{\bf{U}}_n+\sigma^2{\bf{U}}_n=\sigma^2{\bf{U}}_n$, which readily implies ${{\bf{A}}}^H(\boldsymbol{\phi},\boldsymbol{\theta}){\bf{U}}_n=0$, the pseudo spectrum function of MUSIC for the azimuth and elevation angles is
 \begin{equation}
 \label{MUSIC}
 {\bf{P}}(\boldsymbol{\phi},\boldsymbol{\theta}){{ = }}\frac{1}{{{{{{\bf{A}}^H(\boldsymbol{\phi},\boldsymbol{\theta})}{{\bf{U}}_n}{\bf{U}}^H_n{\bf{A}}(\boldsymbol{\phi},\boldsymbol{\theta})^H}}}},
 \end{equation}
where the estimated angles including $\boldsymbol{\hat \phi}$ and $\boldsymbol{\hat \theta}$ can be obtained by finding the $K$ largest peaks of ${\bf{P}}(\boldsymbol{\phi},\boldsymbol{\theta})$. Note that the estimation of azimuth-elevation angles does not need the Doppler frequency and symbol to be known as ${\boldsymbol{\omega}}({\boldsymbol { f}}_{\boldsymbol{D}},l)$ and ${\bf{ s}}_{t,l}$ in ${\bf{R}_W}$ are cancelled out when we obtain ${{\bf{A}}}(\boldsymbol{\phi},\boldsymbol{\theta})^H{\bf{U}}_n=0$. Therefore, no alternative optimisation process is needed in the joint MUSIC-MLE algorithm, which saves the computation complexity.

Afterward, each item of ${\bf{ Y}}_1$ is multiplied by ${\bf{\hat A}}^{-1}={{\bf{A}}}^{-1}({\boldsymbol{\hat \phi}}, {\boldsymbol{\hat \theta}})$ to compensate for $\boldsymbol{ \phi}$ and $\boldsymbol{ \theta}$ in ${\bf{ Y}}_1$ as ${\bf{\hat A}}^{-1}{\bf{ Y}}_1=[{\bf{\hat A}}^{-1}{\bf{  y}}_{0,l},..., {\bf{\hat A}}^{-1}{\bf{ y}}_{t,l}] $. Define a new vector to represent the vector form of ${\bf{\hat A}}^{-1}{\bf{ Y}}_1$ as ${\bf{ y}}_2$, which is ${\bf{ y}}_2=[({{\bf{\hat A}}}^{-1}{{\bf{  y}}_{0,l}})^T,({{\bf{\hat A}}}^{-1}{{\bf{  y}}_{1,l}})^T, ...,({{\bf{\hat A}}}^{-1}{{\bf{ y}}_{t,l}})^T]^T$. Thereafter, the MLE can be applied to estimate ${\boldsymbol f}_{\boldsymbol{D}}$ and detect symbol ${\bf{ \hat s}}_{t,l}$ as 
\begin{equation}
\label{ML2}
 \left[{\bf{ \hat s}}_{t,l}, {\boldsymbol {\hat f}}_{\boldsymbol{D}} \right]  = \mathop {\arg \min }\limits_{{\bf{  s}}_{t,l } , {\boldsymbol f}_{\boldsymbol{D}}} \left|\left|{\bf{ y}}_2 -
\begin{bmatrix}
{\bf{\hat A}}^{-1}{\bf{\hat A}}{\boldsymbol{\omega}}(l){{\bf{s}}_{0,l}}
\\ 
{\bf{\hat A}}^{-1}{\bf{\hat A}}{\boldsymbol{\omega}}(l){{\bf{\hat s}}_{1,l}}\vspace{-0.1cm}\\
\vdots
\vspace{-0.1cm}
\\{\bf{\hat A}}^{-1}{\bf{\hat A}}{\boldsymbol{\omega}}(l){{\bf{s}}_{t,l}}
\end{bmatrix}\right|\right|^2_2.
\end{equation}

The reason why the joint MUSIC-MLE algorithm can obtain an enhanced performance with the increase of the length of the time window is the same as that of the alternating MLE-MLE algorithm. A more detailed description of the joint MUSIC-MLE algorithm can be found in Algorithm 2.

\begin{algorithm}[tp] 
	\caption{The joint MUSIC-MLE algorithm}
	\label{Al2}
  \vspace{0.1cm}
\textbf{\mbox{Time window ($1$) (Pilot signal):}}
 \vspace{-0.4cm}
	\begin{algorithmic}[1]    
       \STATE Calculate $\boldsymbol{\hat \phi}$ and $\boldsymbol{\hat \theta}$
       by using MUSIC in \eqref{MUSIC}, and then calculate ${\boldsymbol{\hat f} }_{\boldsymbol{D}}$ by using ML in \eqref{ML2}. 
       \STATE {{\bf output:}} $\boldsymbol{\hat \phi_0}$, $\boldsymbol{\hat \theta_0}$, ${\boldsymbol{ \hat f}}_{\boldsymbol{D,0}}$.
          	\end{algorithmic}
 \textbf{\mbox{ Time window ($t+1$) (Pilot signal and $t$ symbols):}}
 \vspace{-0.4cm}
	\begin{algorithmic}[1]    
       \STATE Calculate $\boldsymbol{\hat \phi}$ and $\boldsymbol{\hat \theta}$
       by using MUSIC in \eqref{MUSIC}, and then calculate ${\boldsymbol{\hat f} }_{\boldsymbol{D}}$ and ${\bf{ \hat s}}_{t,l}$ by using ML in \eqref{ML2}.
       \STATE {{\bf output:}} $\boldsymbol{\hat \phi_{t}}$, $\boldsymbol{\hat \theta_{t}}$, ${\boldsymbol{ \hat f}}_{\boldsymbol{D,{t}}}$ and ${\bf{  \hat s}}_{\boldsymbol{t,l}}$.
          	\end{algorithmic}
 \textbf{\mbox{ Time window ($T+1$) (Pilot signal and $T$ symbols):}}
 \vspace{-0.4cm}
       \begin{algorithmic}[1]
       \STATE Obtain the final estimation of location parameters and detect the last symbol using the same process as that of time window ($t+1$),
       \STATE {{\bf output:}} $\boldsymbol{\hat \phi_{T}}$, $\boldsymbol{\hat \theta_{T}}$, ${\boldsymbol{ \hat f}}_{\boldsymbol{D,{T}}}$ and ${\bf{  \hat s}}_{\boldsymbol{T,l}}$.
       \end{algorithmic}
\end{algorithm}

\section{Performance analysis for DLDD}
\label{performance analysis DLDD}
In this section, the performance analysis for the scenario DLDD is provided where the average CRLB is derived to evaluate the performance limit of the localisation, while the average SDR is derived for assessing the communication part under the condition of gain-phase defects.

\subsection{ Average Cramér-Rao lower bound (CRLB)}

Based on the MLE in Sec. \ref{locsalition algorithm}-\ref{MLE localisation}, the average CRLB for the azimuth-elevation-Doppler estimation is derived in this section as a benchmark for the localisation performance of the considered PASCAL system with gain-phase errors. In order to obtain the average CRLB, a Fisher information matrix (FIM) is derived first as 
\begin{equation}
{\bf{F}} \triangleq  - \mathbb{E}\Big[{{{\partial ^2}\ln f({{\bf{\bar y}}}|\boldsymbol{\beta} )}}/({{\partial \boldsymbol{\beta} \partial {\boldsymbol{\beta} ^T}}}) \Big],
\end{equation}
where ${\bf{F}}\in \mathbb{C}{^{3K \times 3K}} $ and $\partial f({{\bf{\bar y}}}|\boldsymbol{\beta} )/\partial \boldsymbol{\beta}$ indicates the partial derivative of $f({{\bf{\bar y}}}|\boldsymbol{\beta} )$ with respect to $\boldsymbol{\beta}$. As a consequence, the $(i, j)$th submatrix of ${\bf{F}}$ is written as follows.
\begin{equation}
\label{F_ij}
{\bf{F}}_{i, j} \triangleq  - \mathbb{E}\Big[{{{\partial ^2}\ln f({\bf{\bar y}}|{\boldsymbol{\beta }})}}/({{\partial \boldsymbol{{{\beta }}}_i\partial \boldsymbol{{{\beta }}}}_j})\Big], 
\end{equation}
where ${\bf{F}}_{i, j}\in \mathbb{C}{^{3 \times 3}} $. According to the Slepian-Bangs formula \cite[pp.363]{Slepian-Bangs}, which is specially designed for the multivariate Gaussian case, ${\bf{F}}_{i, j}$ can also be evaluated by 
\begin{equation}
\label{F_ij2}
{{\bf{F}}_{i, j}} = \mathrm{tr}\left[{{\bf{\Gamma }}^{ - 1}}\frac{{\partial {\bf{\Gamma }}}}{{\partial {{\boldsymbol{\beta} }_i}}}{{\bf{\Gamma }}^{ - 1}}\frac{{\partial {\bf{\Gamma }}}}{{\partial {{\boldsymbol{\beta} }_j}}}\right] + 2{{\Re}} \left[\frac{{\partial {\boldsymbol{\mu} ^H}}}{{\partial {{\boldsymbol{\beta} }_i}}}{{\bf{\Gamma }}^{ - 1}}\frac{{\partial \boldsymbol{\mu} }}{{\partial {{\boldsymbol{\beta} }_j}}}\right].
\end{equation}
Since $\displaystyle {{\partial {{\bf{\Gamma }}}}}/{{\partial {{\boldsymbol{\beta }}_i}}}=$ $\displaystyle {{\partial {{\bf{\Gamma }}}}}/{{\partial {{\boldsymbol{\beta }}_j}}}=0$, ${\bf{F}}_{i, j}$ reduces to
\vspace{0.1cm}
\begin{equation}
\label{simplified version}
{\bf{F}}_{i, j} = 2{\mathop \Re \nolimits} \left[{{\partial {\boldsymbol{\mu} ^H}}}/({{\partial {{\boldsymbol{\beta} }_i}}}){{\boldsymbol{\Gamma }}^{ - 1}}{{\partial \boldsymbol{\mu} }}/({{\partial {{\boldsymbol{\beta} }_j}}})\right],
\vspace{0.1cm}
\end{equation}
where the derivation of $\boldsymbol{\mu}$ and ${\boldsymbol{\Gamma }}$ can be found in Sec. \ref{locsalition algorithm}-\ref{MLE localisation}. ${{\boldsymbol{\beta} }_i}\!=\!{[{{\boldsymbol{\phi}}(i) },{{{\boldsymbol{\theta}}}(i) },{{\boldsymbol{f_{D}}}(i) }]^T}$  and ${{\boldsymbol{\beta} }_j}={[{{\boldsymbol{\phi}}(j) },{{{\boldsymbol{\theta}}}(j) },{{\boldsymbol{f_{D}}}(j) }]^T}$. The derivations of $ \displaystyle {{\partial {\boldsymbol{\mu}}}}/{{\partial {{\boldsymbol{\beta} }_i}}}$ and $\displaystyle {{\partial {\boldsymbol{\mu}}}}/{{\partial {{\boldsymbol{\beta} }_j}}} $ can be obtained by using the derivation of their general expression  $\displaystyle {{ {\partial \boldsymbol{\mu}}}}/{{\partial {{\boldsymbol{\beta} }_k}}} = [{{ \partial {\boldsymbol{\mu}}}}/{\partial{\boldsymbol{\psi}} (k)},{\partial{ {\boldsymbol{\mu}}}}/{{\partial {{\boldsymbol{f_{D}}} (k)}}}]^T  $, where ${{\boldsymbol{\psi}} (k)} \in \{{{\boldsymbol{\phi}} (k)},{{\boldsymbol{\theta}} (k)}\}$ and the elements in ${{ {\partial \boldsymbol{\mu}}}}/{{\partial {{\boldsymbol{\beta} }_k}}}$ can be evaluated as
\vspace{0.1cm}
\begin{subequations}
\label{small}
\small
\begin{equation}
\label{muphi0}
\begin{array}{*{20}{l}}
\! \! \! \! \displaystyle \frac{{\partial \boldsymbol{\mu} }}{{\partial \! {{\boldsymbol{\psi}}\! (k)\!}}}\! \!=
\!\!
\sqrt{\!{{P}}_{k}\!}{\Big\{\! \Big[ \frac{{\partial {\mathbb{E}}[{{\bf{ a}}}({\phi _k},{\theta _k})]}}{{\partial {{\boldsymbol{\psi}} (k)}}}{{{\omega}}_k}\!(1)\!\Big]^T\!\!,\!..., \!\Big[\frac{{\partial {\mathbb{E}}[{{\bf{ a}}}({\phi _k},{\theta _k})]}}{{\partial {{\boldsymbol{\psi}} (k)}}}{{{\omega}}_k}\!(L)\!\Big]^T\!\!\Big\} \!}^T,
\end{array}
\end{equation}
\begin{equation}
\label{muf0}
\begin{array}{l}
\! \! \! \!\displaystyle \frac{{\partial \boldsymbol{\mu} }}{{\partial \!{{\boldsymbol{f\!_{D}}} \!(k)\!}}}\!\! = \!\!
\sqrt{\!{{P}}_{k}\!}{ \Big\{\!\Big[\!{\mathbb{E}}[{{\bf{ a}}}({\phi _k},{\theta _k})] \!\frac{{ \partial{{{\omega}}_k}\!(1)\!}}{{\partial {{\boldsymbol{f_{D}}} (k)}}}\!\Big]^T\!\!,\!...,\! \Big[\!{\mathbb{E}}[{{\bf{ a}}}({\phi _k},{\theta _k})]\!\frac{{\partial {{{\omega}}_k}\!(L)\!}}{{\partial {{\boldsymbol{f_{D}}} (k)}}}\Big]^T\!\!\Big\}\!}^T.
\end{array}
\end{equation}
\vspace{0.1cm}
\end{subequations}
where $ {{\boldsymbol{\omega}}_k}(l) = \eta_k{e^{  j2\pi {f_{D,k}}l/{f_s}}}$. In order to obtain \eqref{small}, the fact that $\displaystyle{{\partial {\mathbb{E}}[{{\bf{ a}}}({\phi _{ p}},{\theta _{ p}})]}}/{{\partial {{\boldsymbol{\psi}} (k) }}}\times{{{\omega}}_{p}}(l)=0$ and $\displaystyle{\mathbb{E}}[{{\bf{ a}}}({\phi _{p}},{\theta _{p}})]\times {{ \partial{{{\omega}}_{p}}(l)}}/{{\partial {{\boldsymbol{f_{D}}} (k)}}}=0$ $\forall \ p \ne k$ have been employed.

   \begin{figure*}[!b]
\hrulefill
\label{gammaxgammay}
\setcounter{equation}{39}
\begin{align}
\label{gammaxgammay}
{\mathbb{E}}&[{|{{\boldsymbol{\hat {h } }^H_k}}   {{\boldsymbol{ {h } }_k}}{|^2}}{|{\boldsymbol{\hat {h } }^H_k}}  {{\boldsymbol{ {h } }_p}}|^2]\!\!  =\!\! \!\!\!\!{\sum\limits_{m_1,m_2,m_3,m_4  \in S_1} }{\sum\limits_{n_1,n_2,n_3,n_4  \in S_2} } \!\!\!\! \underbrace{{\mathbb{E}}[{\alpha_{m_1,n_1}}{\alpha_{{ m_2},{ n_2}}}{\alpha_{{ m_3},{ n_3}}}{\alpha_{{ m_4},{ n_4}}}]}_{{\mathbb{E}}_\alpha} \underbrace{\mathbb{E}{[{e^{j(\Delta {\delta _{m_1,n_1}}-\Delta {\delta _{{ m_2},{ n_2}}}+\Delta {\delta _{{ m_3},{ n_3}}}-\Delta {\delta _{{ m_4},{ n_4}}})  }}]}}_{{\mathbb{E}}_\delta} \nonumber \\  
 &\times\!\!
\eta^2_k\eta^2_p {{{   a}_{m_1,n_1}}}\!(\!{ \phi _k},{ \theta _k}\!){{{   a}^*_{m_2,n_2}}}\!(\!{ \phi _k},{ \theta _k}\!){{{   a}_{m_3,n_3}}}\!(\!{ \phi _p},{ \theta _p}\!){{{   a}^*_{m_4,n_4}}}\!(\!{ \phi _p},{ \theta _p}\!) \underbrace{{\mathbb{E}}[{{{   a}^*_{m_1,n_1}}}\!({ \hat \phi _k},{\hat \theta _k}){{{   a}_{m_2,n_2}}}\!({\hat \phi _k},{\hat \theta _k}) {{{   a}^*_{m_3,n_3}}}\!({ \hat \phi _k},{\hat \theta _k}){{{   a}_{m_4,n_4}}}\!({\hat \phi _k},{\hat \theta _k})]}_{{\mathbb{E}}_2}.
\end{align}
\vspace{-0.2cm}
\end{figure*}

Define the $l$-th item in $\displaystyle {{\partial \boldsymbol{\mu} }}/{{\partial {{\boldsymbol{\psi}} (k)}}}$ and $\displaystyle {{\partial \boldsymbol{\mu} }}/{{\partial {{\boldsymbol{f_{D}}} (k)}}}$ as ${\boldsymbol{\Omega} _{{\psi _k}}}$ and ${\boldsymbol{\Omega} _{{f_{D,k}}}}$, respectively, which can be expressed as
\vspace{0.1cm}
\begin{align}
\label{ell}
{\boldsymbol{\Omega} _{{\psi _k}}} &=  {{\bf{\Lambda }}_{{\psi _k}}}\sqrt{{{P}}_{k}}{\mathbb{E}}[{\bf{\tilde a}}({\phi _k},{\theta _k})]{\omega _k}(l), \tag{27a} \\
{\boldsymbol{\Omega} _{{f_{D,k}}}} &= j 2\pi {l}/{f_s}\sqrt{{{P}}_{k}}{\mathbb{E}}[{\bf{\tilde a}}({\phi _k},{\theta _k})]{\omega _k}(l),  \tag{27b} 
\end{align}
\vspace{0.1cm}
where\setcounter{equation}{28} the $\{m,n\}$th element of ${\mathbb{E}}[{{\bf{\tilde a}}}({\phi _k},{\theta _k})]$ can be denoted by using ${\mathbb{E}}[{\alpha _{m,n}}]{\mathbb{E}}[{e^{j\Delta {\delta _{m,n}}}}]{{ {a}}}_{m,n}({\phi _k},{\theta _k})$. 

It is worth mentioning that the closed-form expression of ${\mathbb{E}}[{\alpha _{m,n}}]$ can be obtained by using Theorem 1, and the closed-form expression of ${\mathbb{E}}[{e^{j\Delta {\delta _{m,n}}}}]$ can be obtained using the derivation of its general expression ${\mathbb{E}}\big[{e^{\pm jC\Delta {\delta _{\rho}}}}\big]$ in Appendix A. 

${{\bf{\Lambda }}_{{\psi _k}}}$ in \eqref{ell} indicates the diagonal matrices, which is
\begin{equation}
\label{Lambda}
\begin{array}{l}
\  {{\boldsymbol{\Lambda} _{{\psi _k}}} \triangleq \mathrm{diag}\big[ {\Lambda _{1,1}\,...,{\Lambda _{M,N}}}\big]},  \ \
\end{array}
\end{equation}
where ${\Lambda _{m,n}}\in\{{ \Phi _{m,n}},{ \Theta _{m,n}}  \} $, which can be written as
\begin{subequations}
\label{parameters4}
\begin{equation}
\begin{array}{l}
\!\!\!\!\!{ \Phi _{m,n}}\!\! = \!\! - j2\pi  \{ [-(\!m\!\! -\!\! 1\!){d}\!\sin\! {\phi _k}\!\!+\!\! (n\! \!-\!\! 1){d}\! \cos\! {\phi _k} ]\!\sin\! {\theta _k}\!/\!\lambda, \!\!\!
\end{array}
\end{equation}
\begin{equation}
\begin{array}{l}
\!\!\!\!\!\!{ \Theta _{m,n}}\!\! = \!\! - j2\pi  \{ [(\!m\!\! -\!\! 1\!){d}\!\cos\! {\phi _k}\!\!+\!\! (n\! \!-\!\! 1){d}\! \sin\! {\phi _k} ]\!\cos\! {\theta _k}\!/\!\lambda.\!\!\!
\end{array}
\end{equation}
\end{subequations}

It should be noted that the derived CRLB is more practical than the deterministic CRLBs in \cite{Performance 3} since the latter ones are conditioned on gain-phase errors and they ignore that gain-phase errors vary randomly in the derivation of CRLBs. Nevertheless, the derived CRLB is a function of the average gain-phase errors since both the mean vector $\boldsymbol{\mu}$ and covariance matrix ${\bf{\Gamma }}$ in \eqref{simplified version} contain the expected value of gain-phase defects. It is worth mentioning that $\boldsymbol{\mu}$ and ${\bf{\Gamma }}$ employed in CRLB have been derived in the MLE algorithm in Sec. \ref{locsalition algorithm}-\ref{MLE localisation} by noticing that the derivation of CRLB is corresponding to MLE. In addition, we employ a generalised model to model the statistical characteristics of gain-phase errors, which makes the derived CRLB more complex but more general. 

\subsection{The Average Sum Data Rate (SDR)}
\label{com_A}
In order to evaluate SDR for the communication stage of DLDD in Sec. \ref{locsalition algorithm}-\ref{com stage}, the signal-to-interference-plus-noise ratio (SINR) of the $k$th drone is required, which can be denoted as 
\vspace{-0.2cm}
\begin{equation}
{ \gamma _k} = \frac{{P_k |{{\boldsymbol{\hat {h } }^H_k}} {{\boldsymbol{ {h } }_k}}{|^2}}}{\!\!\!{{\sum\limits_{p = 1,p \ne k}^K \!\!\!P_p{|{\boldsymbol{\hat {h } }^H_k}} {{\boldsymbol{ {h } }_p}}|^2  + |{{\boldsymbol{\hat {h } }^H_k}}{\bf{ n}}{|^2}}}},
\label{gamma_k}
\vspace{-0.2cm}
\end{equation}
where $ {{\boldsymbol{\hat {h } }^H_k}} \! {{\boldsymbol{ {h } }_\varsigma}} \!\!\!=\!\!\!\!\! {\sum\limits_{m = 1}^M }{\sum\limits_{n = 1}^N}  {{{\hat {h } }^*_{k,m,n}}}{{{ {h } }_{\varsigma,m,n}}}$, in which ${{{\hat {h } }_{k,m,n}}}$ and ${{{ {h } }_{\varsigma,m,n}}}$ refer to the $(m,n)$th element of ${\boldsymbol{\hat {h } }_k}$ and ${\boldsymbol{ {h } }_\varsigma} $ for $\varsigma \in\{k,p\}$, and $  {{\boldsymbol{\hat {h } }^H_k}} \! {\bf{ n}}\!\! =\!\!\!\!\! {\sum\limits_{m = 1}^M }{\sum\limits_{n = 1}^N}  {{{\hat {h } }^*_{k,m,n}}} n_ {m,n}$. Note that ${\boldsymbol{\hat {h } }_k}$ is obtained by using the location parameters estimated by the localisation algorithms, which are influenced by gain-phase defects and AWGN noise, and ${\boldsymbol{ {h } }_\varsigma}$ itself is a function of gain-phase errors. Therefore, gain-phase errors can affect the communication performance of our system in two aspects, and thus the performance analysis is challenging.  ${{{\hat {h } }_{k,m,n}}}$ and ${{{ {h } }_{\varsigma,m,n}}}$ can be written as  
\begin{subequations}
    \begin{equation}
    \label{h_n}
\!\!\!\!\!{{{\hat {h } }_{k,m,n}}}\!\! = \!\! {{{   a}_{m,n}}}({\hat \phi _k},{\hat \theta _k}) {{\mathrm{e}}^{j2\pi {\hat f_{D,k}}l/{f_s}}}, 
\vspace{-0.1cm}
\end{equation}
    \begin{equation}
    \label{h_varsigma}
{{{ {h } }_{\varsigma,m,n}}}\!\! =\!\!\eta_k{\alpha _{m,n}}{e^{j\Delta {\delta _{m,n}}}}{{{   a}_{m,n}}}({ \phi _\varsigma},{ \theta _\varsigma}) {{\mathrm{e}}^{j2\pi { f_{D,\varsigma}}l/{f_s}}},
 \end{equation}
\end{subequations}
where  ${{{   a}_{m,n}}}({\hat \phi _k},{\hat \theta _k})\!\!=\!\! {e^{\!- \!j2\pi  [({m}\!-\!1)d\cos\! {\hat \phi _k}\!+\!({n}\!-\!1)d\sin\! {\hat \phi _k}]\sin\! {\hat \theta _k}/\lambda}}\vspace{0.1cm}$ and ${{{   a}_{m,n}}}({ \phi _\varsigma},{ \theta _\varsigma})\!\!=\!\! {e^{\!- \!j2\pi  [({m}\!-\!1)d\cos\! {\phi _\varsigma}\!+\!({n}\!-\!1)d\sin\! { \phi _\varsigma}]\sin\! { \theta _\varsigma}/\lambda}}$. Based on \eqref{gamma_k}, the uplink sum rate can be obtained using $\footnotesize R \!\! = \!\!\sum\limits_{k = 1}^K {{R_k}}$, where ${{R_k}}$ denotes the average data rate for the $k$th user, which can be calculated using ${R_k} = {\mathbb{E}} [ {\log _2}(1 + {{ \gamma }_k})]$, where ${{ \gamma }_k}$ contains multiple random variables including ${\boldsymbol{\hat {h } }_k}$, ${{\boldsymbol{ {h } }_k}} $, ${{\boldsymbol{ {h } }_p}} $ and ${\bf{ n}} $.
 
\subsection{An Approximation Method for the Average SDR} \label{com_B}

Due to the fact that the numerator and denominator of ${{ \gamma }_k}$ are highly correlated, directly calculating the closed-form expression for $R_k$ is not feasible as the distribution of ${{ \gamma }_k}$ is not traceable. In the literature, the first-order Taylor approximation shown below is usually employed to approximate the average data rate \cite{approx1},
\setcounter{equation}{32}
 \begin{equation}
 \label{approximation1}
{\mathbb{E}}[{\log _2}(1 \!+\! {\gamma_x}/{\gamma_y})] \!\approx \! {\log _2}(1 \!+\! {{\mathbb{E}}[\gamma_x]}/{{\mathbb{E}}[\gamma_y]}),
\end{equation}
where $\gamma_x$ and $\gamma_y $ indicate the numerator and denominator of $\gamma_k$. However, a more accurate approximation method is proposed in this paper by using second-order Taylor expansions. First of all, by applying the second-order Taylor approximation to approximate $ {\mathbb{E}}[{\log _2}(1 \!+\! { \gamma _k})]$, we obtain 
 \begin{equation}
 \label{approximation method}
{\mathbb{E}}[{\log _2}(1 \!+\! { \gamma _k})] \!\approx \! {\log _2}(1 \!+\! {\mathbb{E}}[{ \gamma _k}]) - \frac{{{\mathbb{E}}[ \gamma _k^2] - {{{\mathbb{E}}[{{ \gamma }_k}]}^2}}}{{2\ln (2){{(1 + {\mathbb{E}}[{{ \gamma }_k}])}^2}}},
\end{equation}
Nevertheless, since ${\mathbb{E}}[{{ \gamma }_k}]$ and ${\mathbb{E}}[ \gamma _k^2]$ are intractable, the second-order Taylor approximation is employed again to obtain approximations for them, which are denoted by
 \vspace{-0.2cm}
\begin{subequations}
\label{E_gamma_k_k2}
 \begin{equation}
{\mathbb{E}}[{ \gamma _k}]  \!\!\approx\!\! \frac{{{\mathbb{E}}[{{{\gamma}} _x}]}}{{{\mathbb{E}}[{{{\gamma}} _y}]}} - \frac{{{\mathop{\rm cov}} ({{{\gamma}} _x},{{{\gamma}} _y})}}{{{{{\mathbb{E}}[{{{\gamma}} _y}]}^2}}} + \frac{{{\mathop{\rm var}} ({{{\gamma}} _y}){\mathbb{E}}[{{{\gamma}} _x}]}}{{{{{\mathbb{E}}[{{{\gamma}} _y}]}^3}}},
\end{equation}
 \begin{equation}
\!\!{\mathbb{E}}[ \gamma _k^2] \!\!\approx\!\! \frac{{{{{\mathbb{E}}[{{{\gamma}} _x}]}^2}}}{{{{{\mathbb{E}}[{{{\gamma}} _y}]}^2}}} + \frac{{{\mathop{\rm var}} (\!{{{\gamma}} _x}\!)}}{{{{{\mathbb{E}}[{{{\gamma}} _y}]}^2}}} - \frac{{\!4{\mathbb{E}}[{{{\gamma}} _x}]{\mathop{\rm cov}} (\!{{{\gamma}} _x},{{{\gamma}} _y}\!)}}{{{{{\mathbb{E}}[{{{\gamma}} _y}]}^3}}} + \frac{{\!3{{{\mathbb{E}}[{{{\gamma}} _x}]}^2}\!{\mathop{\rm var}} (\!{{{\gamma}} _y}\!)}}{{{{{\mathbb{E}}[{{{\gamma}} _y}]}^4}}}\!,
\vspace{-0.1cm}
\end{equation}  
\end{subequations}
where ${{\mathop{\rm var}} ({{{\gamma}} _x})}$, ${\mathop{\rm var}} ({{{\gamma}} _y})$ and ${\mathop{\rm cov}} ({{{\gamma}} _x},{{{\gamma}} _y}) $ can be calculated using their definitions, i.e., ${\rm{var}}({{\gamma}}) = {\mathbb{E}}[{{\gamma^2}}] - {\mathbb{E}}[{{{\gamma}}}]^2$ and ${\mathop{\rm cov}} ({{{\gamma}} _x},{{{\gamma}} _y}) = {\mathbb{E}}[{{{\gamma}} _x} {{{\gamma}} _y}] - {\mathbb{E}}[{{{\gamma}} _x}]{\mathbb{E}}[{{{\gamma}} _y}]$. 

To begin, we want to calculate ${\mathbb{E}}[{{{\gamma}} _x}] $ and ${\mathbb{E}}[{{{\gamma}} _y}] $. By utilising some mathematical manipulations and the fact that the variables including estimated azimuth angle, estimated elevation angle, estimated Doppler frequency, gain errors, phase defects are all independent of each other, ${\mathbb{E}}[{{{\gamma}} _x}]$ can be derived. The expression of ${\mathbb{E}}[{{{\gamma}} _x}]$ can be obtained by substituting $\varsigma=k $ into its general expression ${\mathbb{E}}[{{{\gamma}} _\varsigma}]=P_\varsigma {\mathbb{E}}[|{{\boldsymbol{\hat {h } }^H_{k}}} {{\boldsymbol{ {h } }_{\varsigma}}}{|^2}]$ for $\varsigma \in\{k,p\}$, where ${\mathbb{E}}[|{{\boldsymbol{\hat {h } }^H_{k}}} {{\boldsymbol{ {h } }_{\varsigma}}}{|^2}]$ is shown in \eqref{gammax} on bottom of page \pageref{gammax}, in which $S_1=\{1,...,M\}$ and $S_2=\{1,...,N\}$. The derivations of ${\mathbb{E}}[{\alpha _{m_1,n_1}}{\alpha _{m_2,n_2}}]$ and ${\mathbb{E}}[{e^{j(\Delta {\delta _{m_1,n_1}} - \Delta {\delta _{m_2,n_2}})}}]$ can be found in Sec. \ref{locsalition algorithm}-\ref{MLE localisation}, and the detailed derivation of ${\mathbb{E}}_1$ is included in Appendix \ref{Appendix B}. ${\mathbb{E}}[{{{\gamma}} _y}]$ can be calculated by using the linearity of expectation as 
\setcounter{equation}{36}
\begin{equation}
\label{gammary}
\begin{array}{l}
 {\mathbb{E}}[{{{\gamma}} _y}] \!\! =
\!\! {{\sum\limits_{p = 1,p \ne k}^K P_p{\mathbb{E}}[{|{\boldsymbol{\hat {h } }^H_k}} {{\boldsymbol{ {h } }_p}}|^2]  + {\mathbb{E}}[|{{\boldsymbol{\hat {h } }^H_k}}{\bf{ n}}{|^2}}}], 
\end{array}
\end{equation}
where ${\mathbb{E}}[{|{\boldsymbol{\hat {h } }^H_k}} {{\boldsymbol{ {h } }_p}}|^2]$ can be obtained by using the general expression ${\mathbb{E}}[|{{\boldsymbol{\hat {h } }^H_{k}}} {{\boldsymbol{ {h } }_{\varsigma}}}{|^2}]$ shown in \eqref{gammax} on bottom of page \pageref{gammax}. ${\mathbb{E}}[|{\boldsymbol{\hat {h } }^H_k}{\bf{ n}}{|^2}]$ can be denoted by
\begin{equation}
\label{gammarn}
{\mathbb{E}}[|{\boldsymbol{\hat {h } }^H_k}{\bf{ n}}{|^2}]=  {\sum\limits_{m_1,m_2 \in S_1} }{\sum\limits_{n_1,n_2 \in S_2} }  {\mathbb{E}}[{ n_{m_1,n_1}}{ n_{{ m_2},{ n_2}}^ *}] {\mathbb{E}}_{ 1},   
\end{equation}

   \begin{figure*}[!b]
\hrulefill
\setcounter{equation}{41}
\begin{align}
\label{gammaxgammay2}
\!\! {\mathbb{E}}[{|{{\boldsymbol{\hat {h } }^H_k}}  \!{{\boldsymbol{ {h } }_\varsigma}}{|^2}}|{{\boldsymbol{\hat {h } }^H_k}} \!{\bf{ n}}{|^2}]\!\! 
=\!\! \!\!\!\!\!\! \!\!\!\!{\sum\limits_{m_1\!,m_2\!,m_3\!,m_4 \! \in S_1} }{\sum\limits_{n_1\!,n_2\!,n_3\!,n_4 \! \in S_2} } \!\!\!\!\!\!\!\!\!\!\eta^2_\varsigma  {\mathbb{E}}[{\alpha_{m_1,n_1}}\!{\alpha_{{ m_2},{ n_2}}}]   {\mathbb{E}}[{e^{j\!(\!\Delta {\delta _{m_1,n_1}}\!-\!\Delta {\delta _{{ m_2},{ n_2}}\!)\!}}}]{\mathbb{E}}[{ n_{{m_3},{n_3}}}{ n_{{ m_4},{n_4}}^*}]{{{   a}_{m_1,n_1}}}\!(\!{ \phi _\varsigma},{ \theta _\varsigma}\!)\! 
{{{   a}^*_{m_2,n_2}}}\!({ \phi _\varsigma},{ \theta _\varsigma}){\mathbb{E}}_2,
\end{align} 
\end{figure*}

It is worth noting that when $m_1 \ne m_2 $ or $n_1 \ne n_2 $, ${\mathbb{E}}[{ n_{{m_1},{n_1}}}{ n_{{m_2},{n_2}}^ *}]= {\mathbb{E}}[{ n_{{m_1},{n_1}}}]{\mathbb{E}}[{ n_{{m_2},{n_2}}^ *}]=0$ as the AWGN corresponding to different antennas is independent. When $m_1 = m_2 $ or $n_1 = n_2 $, ${\mathbb{E}}[{ n_{{m_1},{n_1}}}{ n_{{m_2},{n_2}}^ *}]= \sigma^2_{m.n}$. 

    \begin{figure*}[t]
\centering 

{\includegraphics  [height=1.7in, width=7.1in]{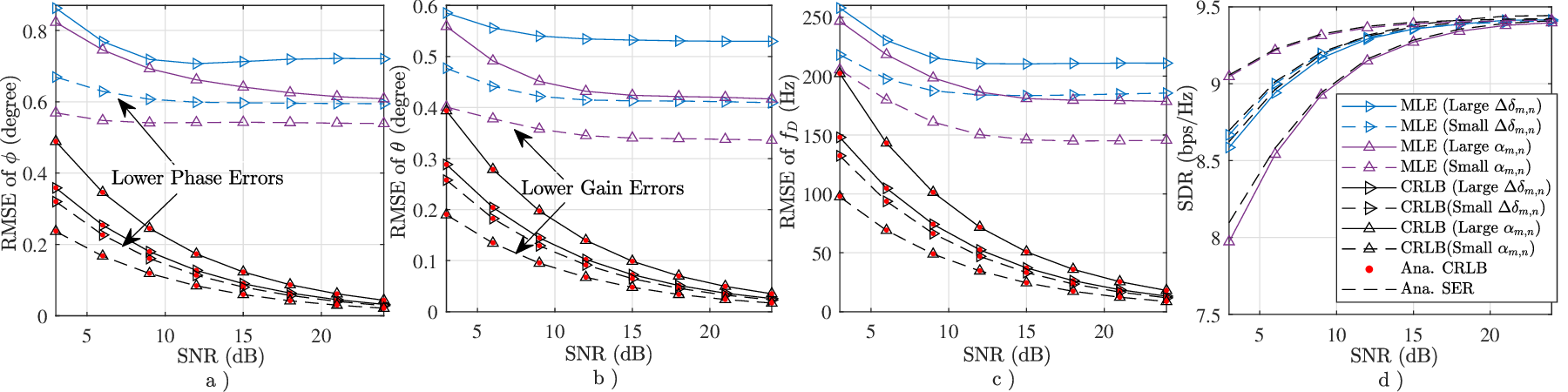}}
 \caption{RMSEs of the a) Azimuth angle $\phi$, b) Elevation angle $\theta$, c) $f_D$ estimations, d) SDR with various gain-phase errors.} 
\label{errorseffect}
\end{figure*}

 \begin{figure}[ptb]
\centering 
{\includegraphics  [height=1.7in, width=3in]{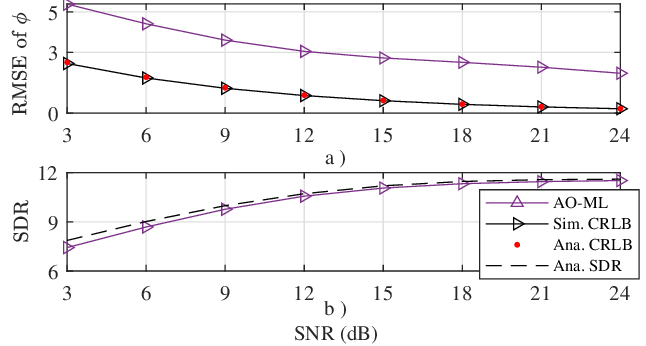}}

\caption{The effect of larger gain-phase errors.}

\label{larger errors}
\end{figure}

 \begin{figure*}[t]
\centering 

{\includegraphics  [height=1.5in, width=7in]{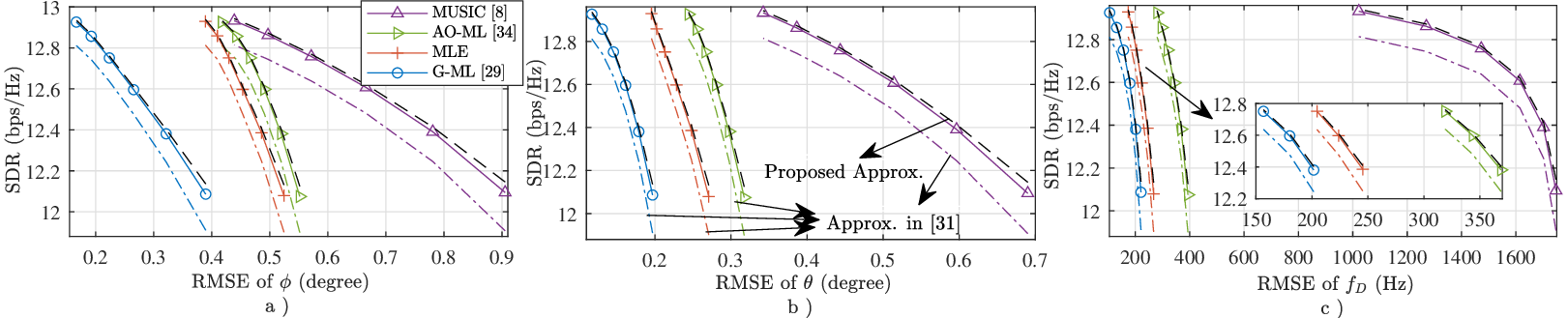}}

\caption{The SDRs with RMSEs of a) Azimuth angle $\phi$, b) Elevation angle $\theta$, c) Doppler frequency $f_D$} 
\label{tradeoff}
\end{figure*}

In order to calculate ${\mathop{\rm cov}} ({{{\gamma}} _x},{{{\gamma}} _y})$ in \eqref{E_gamma_k_k2}, ${\mathbb{E}}[{{{\gamma}} _x} {{{\gamma}} _y}]$ needs to be derived. By using the the linearity of expectation, ${\mathbb{E}}[{{{\gamma}} _x} {{{\gamma}} _y}]$ can be obtained as 
\setcounter{equation}{38}
\begin{equation}
\label{gamma _xy}
\!\!{\mathbb{E}}[{{{\gamma}} _x} {{{\gamma}} _y}] \! \!=\!\! \! \!\! \! \! \!{{\sum\limits_{p = 1,p \ne k}^K \!\!\!\! \!P_k P_p {\mathbb{E}}[{|{{\boldsymbol{\hat {h } }^H_k}}   \!{{\boldsymbol{ {h } }_k}}{|^2}}{|{\boldsymbol{\hat {h } }^H_k}} \! {{\boldsymbol{ {h } }_p}}|^2]  \!\!+\! \!P_k {\mathbb{E}}[{|{{\boldsymbol{\hat {h } }^H_k}}  \!{{\boldsymbol{ {h } }_k}}{|^2}}|{{\boldsymbol{\hat {h } }^H_k}} \!{\bf{ n}}{|^2}]}},\!\!
\end{equation}
where ${\mathbb{E}}[{|{{\boldsymbol{\hat {h } }^H_k}}   \!{{\boldsymbol{ {h } }_k}}{|^2}}{|{\boldsymbol{\hat {h } }^H_k}} \! {{\boldsymbol{ {h } }_p}}|^2]$ can be found in \eqref{gammaxgammay} on bottom of page \pageref{gammaxgammay}. To obtain its closed-form expression, we should consider whether $m_1$, $m_2$, $m_3 $ and $m_4 $ are equal to each other, and whether $n_1$, $n_2$, $n_3 $ and $n_4 $ are equal to each other. For instance, if $m_1=m_2, n_1=n_2$ and $m_2\ne m_3 \ne m_4, n_2\ne n_3 \ne n_4 $,
\begin{equation}
\setcounter{equation}{41}
    {\mathbb{E}}_\alpha={\mathbb{E}}[{\alpha^2_{m,n}}]{\mathbb{E}}[{\alpha_{{ m_3},{ n_3}}}]{\mathbb{E}}[{\alpha_{{ m_4},{ n_4}}}],
\end{equation}
where ${\mathbb{E}}[{\alpha^2_{m,n}}]$, ${\mathbb{E}}[{\alpha_{{ m_3},{ n_3}}}] $ and $ {\mathbb{E}}[{\alpha_{{ m_4},{ n_4}}}]$ can be derived by using Theorem 1 on page \pageref{theorem 1}. To obtain the closed-form expression of ${\mathbb{E}}[{|{{\boldsymbol{\hat {h } }^H_k}}   \!{{\boldsymbol{ {h } }_k}}{|^2}}{|{\boldsymbol{\hat {h } }^H_k}} \! {{\boldsymbol{ {h } }_p}}|^2]$, we also need to calculate ${\mathbb{E}}_{\delta}$ and ${\mathbb{E}}_{ 2} $.
The derivation of ${\mathbb{E}}_{\delta}$ is similar that of ${\mathbb{E}}_\alpha$, and the complete derivation of the closed-from expression of ${\mathbb{E}}_{ 2} $ can be found in Appendix \ref{Appendix B}.

 \begin{figure*}[!b]
\hrulefill
\label{h4}
\setcounter{equation}{43}
\begin{align}
\label{h4}
\!\!\!\!\!\!\!\!\!{\mathbb{E}}[{|{{\boldsymbol{\hat {h } }^H_k}} {{\boldsymbol{ {h } }_\varsigma}}{|^4}}]\!\!  
&=\!\! \!\!{\sum\limits_{m_1,m_2,m_3,m_4  \in S_1} }{\sum\limits_{n_1,n_2,n_3,n_4  \in S_2} } \!\!\!\! \eta^4_\varsigma {{{   a}_{m_1,n_1}}}({ \phi _\varsigma},{ \theta _\varsigma}) 
{{{   a}^*_{m_2,n_2}}}({ \phi _\varsigma},{ \theta _\varsigma})
{{{   a}_{m_3,n_3}}}({ \phi _\varsigma},{ \theta _\varsigma}) 
{{{   a}^*_{m_4,n_4}}}({ \phi _\varsigma},{ \theta _\varsigma}){\mathbb{E}}_\alpha{\mathbb{E}}_\delta{\mathbb{E}}_2,
\end{align}
\setcounter{equation}{46}
\begin{align}
\label{h2h2}
{\mathbb{E}}\left[{|{\bf{\hat {\boldsymbol{h}} }}_k^H{{\bf{{\boldsymbol{h}} }}_p}{|^2}|{\bf{\hat {\boldsymbol{h}} }}_k^H{{\bf{{\boldsymbol{h}} }}_{\tilde p}}{|^2}}\right]\!\!  
&=\!\! \!\!{\sum\limits_{m_1,m_2,m_3,m_4  \in S_1} }{\sum\limits_{n_1,n_2,n_3,n_4  \in S_2} } \!\!\!\!\eta^2_p \eta^2_{\tilde p}  {{{   a}_{m_1,n_1}}}({ \phi _p},{ \theta _p}) 
{{{   a}^*_{m_2,n_2}}}({ \phi _p},{ \theta _p})
{{{   a}_{m_3,n_3}}}({ \phi _{\tilde p}},{ \theta _{\tilde p}}) 
{{{   a}^*_{m_4,n_4}}}({ \phi _{\tilde p}},{ \theta _{\tilde p}}){\mathbb{E}}_\alpha{\mathbb{E}}_\delta{\mathbb{E}}_2,
\end{align}
\end{figure*}

 $ {\mathbb{E}}[{|{{\boldsymbol{\hat {h } }^H_k}}  \!{{\boldsymbol{ {h } }_k}}{|^2}}|{{\boldsymbol{\hat {h } }^H_k}} \!{\bf{ n}}{|^2}]$ in \eqref{gamma _xy} can be obtained by substituting $\varsigma=k $ into the general expression ${\mathbb{E}}[{|{{\boldsymbol{\hat {h } }^H_k}}  \!{{\boldsymbol{ {h } }_\varsigma}}{|^2}}|{{\boldsymbol{\hat {h } }^H_k}} \!{\bf{ n}}{|^2}]$ given in \eqref{gammaxgammay2} on the bottom of page \pageref{gammaxgammay2} for $\varsigma \in\{k,p\}$, where the derivation of ${\mathbb{E}}[{ n_{{m_3},{n_3}}}{ n_{{m_4},{n_4}}^ *}]$ is similar to that of ${\mathbb{E}}[{ n_{{m_1},{n_1}}}{ n_{{m_2},{n_2}}^ *}]$.

To compute ${\rm{var}}({{\gamma_x}})$ in \eqref{E_gamma_k_k2}, we need the value of ${\mathbb{E}}[{\boldsymbol{\gamma}} _x^2]$, which is 
\setcounter{equation}{42}
\begin{equation}
\label{gamma _x^2}
    {\mathbb{E}}[{\boldsymbol{\gamma}} _x^2]=P^2_k {\mathbb{E}}[{|{{\boldsymbol{\hat {h } }^H_k}} {{\boldsymbol{ {h } }_k}}{|^4}}],
\end{equation}
where ${\mathbb{E}}[{|{{\boldsymbol{\hat {h } }^H_k}} {{\boldsymbol{ {h } }_k}}{|^4}}]$ can be derived by performing some  mathematical
manipulations. Afterward, the expression of ${\mathbb{E}}[{|{{\boldsymbol{\hat {h } }^H_k}} {{\boldsymbol{ {h } }_k}}{|^4}}]$ can be obtained by substituting $\varsigma=k $ into the general expression ${\mathbb{E}}[{|{{\boldsymbol{\hat {h } }^H_k}} {{\boldsymbol{ {h } }_\varsigma}}{|^4}}]$ shown in \eqref{h4} on the bottom of page \pageref{h4} for $\varsigma \in\{k,p\}$. 

\begin{table}[t]
\label{R_1_9b}
\centering{}
{\fontsize{8.8pt}{12pt}\selectfont 
\caption{Simulation Parameters}
\newsavebox{\tablebox}
\begin{lrbox}{\tablebox}
\begin{tabular}{|c|c|c|c|c|c|}
\hline
\textbf{Param.}                    &   \textbf{Values} &\textbf{Param.}                    &   \textbf{Values} &\textbf{Param.}                    &   \textbf{Values}
\\
\hline
${\phi}_1$     & $20^ \circ$  & ${\theta}_1$     & $20^ \circ$  & ${f}_{D,1}$     & $ 2\mathrm{kHz}$ 
\\
\hline
${\phi}_2$     & $40^ \circ$  & ${\theta}_2$     & $40^ \circ$  & ${f}_{D,2}$     & $ 4\mathrm{kHz}$ 
\\
\hline
${\phi}_3$     & $60^ \circ$  & ${\theta}_3$     & $60^ \circ$  & ${f}_{D,3}$     & $ 6\mathrm{kHz}$ 
\\
\hline
$v_1$     & $3.4 \; \mathrm{m/s}$  & $v_2$     & $8.4 \; \mathrm{m/s}$  & $v_3$     & $ 19 \; \mathrm{m/s}$
\\
\hline
$T$      &100    &   $f_s$       & 100 $\mathrm{kHz}$ &$\lambda$     & 1.6  $\mathrm{mm}$\\
\hline
\end{tabular}
\end{lrbox}
\scalebox{0.9}{\usebox{\tablebox}} 
\label{sim par}}
\end{table}
  
For the purpose of calculating ${\rm{var}}({{\gamma_y}})$ in \eqref{E_gamma_k_k2}, the derivation of ${\mathbb{E}}[{\boldsymbol{\gamma}} _y^2]$ is required, which can be obtained as 
\setcounter{equation}{44}
\begin{equation}
\label{gamma _y^2}
{\mathbb{E}}[{\boldsymbol{\gamma}} _y^2] = {\mathbb{E}} [{\boldsymbol{\gamma}} _{y1}^2] + 2{\mathbb{E}} [{{\boldsymbol{\gamma}} _{y1}}{{\boldsymbol{\gamma}} _{y2}}] + {\mathbb{E}} [{\boldsymbol{\gamma}} _{y2}^2]  ,
\end{equation}
where ${\boldsymbol{\gamma}} _{y1}=\sum\limits_{p = 1,p \ne k}^K P_p {|{\boldsymbol{\hat {h } }^H_k}} {{\boldsymbol{ {h } }_p}}|^2 $ and ${\boldsymbol{\gamma}} _{y2}=|{{\boldsymbol{\hat {h } }^H_k}}{\bf{ n}}{|^2} $ denote the first and second items of  ${\boldsymbol{\gamma}} _{y}$, in which $\gamma_y $ is the denominator of $\gamma_k$ in \eqref{gamma_k}. By performing some algebraic manipulations, ${\mathbb{E}} [{\boldsymbol{\gamma}} _{y2}^2] $, ${\mathbb{E}} [{{\boldsymbol{\gamma}} _{y1}}{{\boldsymbol{\gamma}} _{y2}}] $ and ${\mathbb{E}} [{\boldsymbol{\gamma}} _{y2}^2] $ can be respectively calculated by 
\begin{subequations}
\vspace{-0.1cm}
\begin{equation}
{\mathbb{E}}[{\boldsymbol{\gamma}} _{y2}^2]\!\!=\!\! {\mathbb{E}}\big[{|{{\boldsymbol{\hat {{\boldsymbol{h}} } }^H_k}}{\bf{ n}}{|^4}}\big], 
\vspace{-0.1cm}
\end{equation}
\begin{equation}
\vspace{-0.1cm}
{\mathbb{E}}[{{\boldsymbol{\gamma}} _{y1}}{{\boldsymbol{\gamma}} _{y2}}]\!\!=\!\! \sum\limits_{p = 1,p \ne k}^K \!\!\!\!\!\! P_p {\mathbb{E}}\big[{|{{\hat {\boldsymbol{h}} }}_k^H} {{\bf{{\boldsymbol{h}} }}_p}{|^2}|{{\hat {\boldsymbol{h}} }}_k^H{\bf{ n}}{|^2}\big], 
\vspace{-0.1cm}
\end{equation}
 \begin{equation}
{\mathbb{E}}[{\boldsymbol{\gamma}} _{y1}^2]\!\!=\!\!\!\!\!\!  \! \!\sum\limits_{p = 1,p \ne k}^K\!\!\!\!\!\!  P^2_p {\mathbb{E}}\big[{|{\boldsymbol{\hat {h } }^H_k}} {{\boldsymbol{ {h } }_p}}|^4\big]+2  \!\!\!\!\!\! \!\!\!\! \sum\limits_{p < \tilde p,p \ne k,\tilde p \ne k}^K \!\!\!\! \!\!\! \!\!\!\! P_p P_{\tilde p} {\mathbb{E}}\big[{|{{\hat {\boldsymbol{h}} }}_k^H{{{{\boldsymbol{h}} }}_p}{|^2}|{{\hat {\boldsymbol{h}} }}_k^H{{{{\boldsymbol{h}} }}_{\tilde p}}{|^2}}\big], 
\vspace{-0.1cm}
\end{equation} 
\end{subequations}
where the derivation of $ {\mathbb{E}}\!\left[{|{\bf{\hat {\boldsymbol{h}} }}_k^H} {{\bf{{\boldsymbol{h}} }}_p}{|^2}|{\bf{\hat {\boldsymbol{h}} }}_k^H{\bf{ n}}{|^2}\right]$ can be obtained by substituting $\varsigma=p $ into its general expression ${\mathbb{E}}[{|{{\boldsymbol{\hat {h } }^H_k}}  \!{{\boldsymbol{ {h } }_\varsigma}}{|^2}}|{{\boldsymbol{\hat {h } }^H_k}} \!{\bf{ n}}{|^2}]$ provided in \eqref{gammaxgammay2} on the bottom of page \pageref{gammaxgammay2}, while the expression of ${\mathbb{E}}\!\big[({|{\boldsymbol{\hat {h } }^H_k}} {{\boldsymbol{ {h } }_p}}|^4)\big]$ can be obtained by substituting $\varsigma=p $ into the general expression ${\mathbb{E}}[{|{{\boldsymbol{\hat {h } }^H_k}} {{\boldsymbol{ {h } }_\varsigma}}{|^4}}]$ shown in \eqref{h4} on the bottom of page \pageref{h4} for $\varsigma \in\{k,p\}$. For the derivations of ${\mathbb{E}}\!\left[{|{\bf{\hat {\boldsymbol{h}} }}_k^H{{\bf{{\boldsymbol{h}} }}_p}{|^2}|{\bf{\hat {\boldsymbol{h}} }}_k^H{{\bf{{\boldsymbol{h}} }}_{\tilde p}}{|^2}}\right]$, its expression can be found in \eqref{h2h2} on the bottom of page \pageref{h2h2}. ${\mathbb{E}}\big[{|{{\boldsymbol{\hat {{\boldsymbol{h}} } }^H_k}}{\bf{ n}}{|^4}}\big]$ can be represented by
\begin{equation}
\setcounter{equation}{48}
  {\mathbb{E}}\big[{|{{\boldsymbol{\hat {{\boldsymbol{h}} } }^H_k}}{\bf{ n}}{|^4}}\big]= {\mathbb{E}}[{ n_{{m_1},{n_1}}}{ n_{{ m_2},{ n_2}}^*}{ n_{{m_{3}},{n_{3}}}}{ n_{{ m_{4}},{ n_{4}}}^*}]{\mathbb{E}}_2,
\end{equation}
whose derivation is similar to that of ${\mathbb{E}}[|{\boldsymbol{\hat {h } }^H_k}{\bf{ n}}{|^2}]$ in \eqref{gammarn}.


In the end, the calculated expected values need to be substituted back to \eqref{approximation method} and \eqref{E_gamma_k_k2} to obtain the average sum data rate  ${\mathbb{E}}[{\log _2}(1 + { \gamma _k})]$. To begin, the calculated results of ${\mathbb{E}}[{{{\gamma}} _x}]$ above \eqref{gammary}, $[{{\boldsymbol{\gamma}} _y}]$ in \eqref{gammary}, ${\mathbb{E}}[{{\boldsymbol{\gamma}} _x} {{\boldsymbol{\gamma}} _y}] $ in \eqref{gamma _xy}, ${\mathbb{E}}[{\boldsymbol{\gamma}} _x^2]$ in \eqref{gamma _x^2}, ${\mathbb{E}}[{\boldsymbol{\gamma}} _y^2]$ in \eqref{gamma _y^2} are substituted into \eqref{E_gamma_k_k2} to calculate ${\rm{var}}({{\gamma_x}}) $, ${\rm{var}}({{\gamma_y}}) $ and ${\mathop{\rm cov}} ({{{\gamma}} _x},{{{\gamma}} _y})$ and then calculate ${\mathbb{E}}[{ \gamma _k}]$ and $ {\mathbb{E}}[ \gamma _k^2] $. Secondly, by substituting the obtained results into \eqref{approximation method}, the approximation value of $ {\mathbb{E}}[{\log _2}(1 + { \gamma _k})]$ can be obtained. Note that there is no ${{\hat f}_{D,k}}$ in ${\mathbb{E}}[{{\boldsymbol{\gamma}} _x}]$, ${\mathbb{E}}[{{\boldsymbol{\gamma}} _y}]$, ${\mathbb{E}}[{{\boldsymbol{\gamma}} _x} {{\boldsymbol{\gamma}} _y}]$, ${\mathbb{E}}[{\boldsymbol{\gamma}} _x^2]$ and ${\mathbb{E}}[{\boldsymbol{\gamma}} _y^2]$ as ${{\hat f}_{D,k}}$ is cancelled out in ${|{{\boldsymbol{\hat {h } }^H_k}} {{\boldsymbol{ {h } }_k}}{|^2}}$ and $|{\boldsymbol{\hat {h } }^H_k} {{\boldsymbol{ {h } }_p}}|^2 $ in \eqref{gamma_k}. Thus, the average sum data rate  ${\mathbb{E}}[{\log _2}(1 + { \gamma _k})]$ is not affected by the estimation errors of Doppler frequency.

\section{Numerical Results}
The simulation and analytical results are provided first to evaluate the performance of DLDD with gain-phase defects, where the root mean squared error (RMSE) and SDR are employed to evaluate the localisation accuracy and communication performance. Afterwards, the performance of the alternating MLE-MLE algorithm and the joint MUSIC-MLE algorithm for JLDD is provided, where RMSE and symbol error rate (SER) are respectively utilised to evaluate the localisation and decoding performance. The simulation parameters can be found in Table \ref{sim par}, where param. denotes parameter. For each simulation point, a number of $10^3$ Monte Carlo tests are performed. In addition, DLDD with $l\!=\!1$ is considered in Figs. \ref{errorseffect}-\ref{larger errors}, while DLDD with $l\!=\!5$ is employed in Fig. \ref{tradeoff} for $l \leq L$. DLDD with $l\!=\!1$ and DLDD with $l\!=\!5$ indicate the data symbols in Fig. \ref{ISAC diagram} are decoded by using the estimated location parameters with  $l\!=\!1$ pilot and $l\!=\!5$ pilots. 

Fig. \ref{errorseffect} shows the effect of different gain-phase defects, on DLDD of the PASCAL system, where four cases of gain-phase defects are considered. In the simulation, a BS with $M\times N=8\times 8$ antennas is employed to localise and serve drone 2 and drone 3, and the number of pilots is
$L \!= \!10$ in the whole frame. In Fig. \ref{errorseffect}, large $\alpha_{m,n}$ and small $\alpha_{m,n}$ respectively refer to the case $\nu=0.5, \sigma_r=0.1, \tilde k=10$ and the case $\nu=1.15, \sigma_r=0.1, \tilde k=10$. Note that the value of gain defects $\alpha_{m,n}$ is controlled by $\nu$ and $\sigma_r$, while the value of phase errors $\Delta \delta_{m,n}$ is controlled by $\tilde k $. By fixing $\sigma_r$ and $\tilde k$, and then varying $\nu$, the value of gain errors will be different. Thus, the effect of gain errors can be explored. Large $\Delta \delta_{m,n}$ and small $\Delta \delta_{m,n}$ in Fig. \ref{errorseffect} respectively denote the case $\nu=0.8, \sigma_r=0.09, \tilde k=5$ and the case $\nu=0.8, \sigma_r=0.09, \tilde k=10$. Since $\nu$ and $\sigma_r$ are fixed in these two cases, we can explore the effect of phase errors. As shown in Fig. \ref{errorseffect}, both the localisation and communication performance of MLE become better in the small $\alpha_{m,n}$ case compared to the large $\alpha_{m,n}$ case as their RMSEs reduce, while their SDRs increases. This is because the perfect case for $\nu$ is 1. Since $\nu$ in small $\alpha_{m,n}$ case is closer to 1, the effect of gain errors will be smaller. Similarly, both the localisation and communication performance of MLE become better in the small $\Delta \delta_{m,n}$ case compared to the large $\Delta \delta_{m,n}$ case since larger $\tilde k $ indicates smaller phase errors in the von Mises distribution when $\tilde \mu=0$. In addition, Fig. \ref{errorseffect} also shows a perfect match between simulated and derived CRLBs and SDRs, which indicates the validity of the derivations in this paper.

In Fig. \ref{larger errors}, the effect of larger gain-phase errors is demonstrated. In the simulation, we consider $\nu=0.5, \sigma_r=0.1$, $ \tilde k=5$, and employ a BS with $M\times N=7\times 7$ antennas at $\mathrm{SNR}=20 \;\mathrm{dB}$. The result indicates that analytical and simulation results for both CRLB and SDR still match very well even if larger gain-phase errors are considered.

Fig. \ref{tradeoff} shows the synergy or win-win relationship between the localisation and communication parts of the considered PASCAL system by deploying drone 1 and drone 3 with a BS composed of $M\times N=8\times 8$ antennas. To obtain these curves, SNR has been changed from $5$ dB to $15$ dB. The parameters for gain-phase errors are set to $\nu\!=\!1, \sigma_r\!=\!0.1$, $ \tilde k\!=\!50$, and $L\!=\!5$ is considered. As shown in Fig. \ref{tradeoff}, smaller RMSEs correspond to larger SDRs for all algorithms, which indicates better localisation results can improve communication performance. Thus the PASCAL system achieves a win-win situation. This is because the estimated location parameters are employed to infer the communication channel information, and thus a more precise channel information can increase SDR. In addition to MLE, Fig. \ref{tradeoff} also demonstrates the performance of the benchmark algorithms including AO-ML in \cite{WCNChan}, MUSIC in \cite{MUSIC1} and genie maximum likelihood (G-ML) in \cite{aerohan}. Interestingly, it can be found that the analytical results for the SDRs corresponding to all localisation algorithms and the simulation results match very well, which indicates that our derivation for the average SDR is general for all localisation algorithms.  This is because the estimated location parameters of all the above-mentioned algorithms follow Gaussian distribution. Fig. \ref{tradeoff} also indicates that the analytical results obtained by using the proposed approximation method in \eqref{approximation method} and the simulation results match very well, while the analytical results by using the first-order approximation method in \cite{approx1} demonstrate a significant discrepancy compared to the simulation results.  This result indicates the superiority of the proposed approximation method in \eqref{approximation method} relative to the first-order approximation method. 

    \begin{figure*}[t]
\centering 

{\includegraphics  [height=2.4in, width=7in]{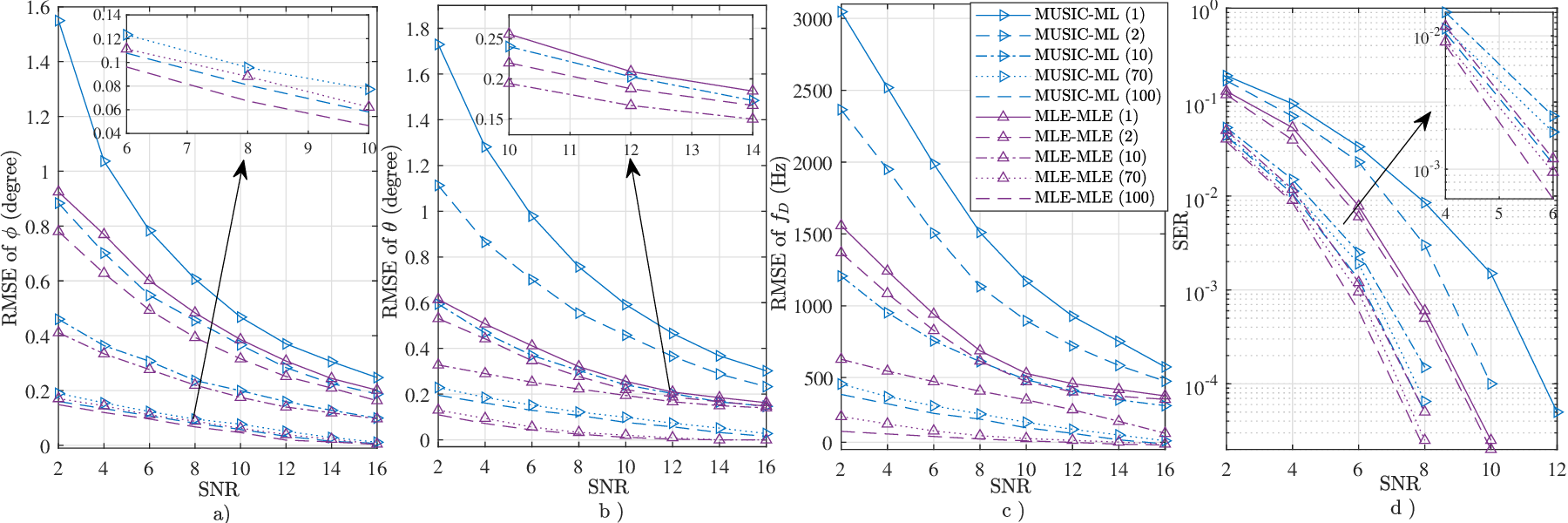}}
 \caption{The performance of the alternating MLE-MLE and joint MUSIC-MLE algorithms.} 
\label{newalgorithms}
\end{figure*}

\begin{figure*}[!b]
\hrulefill
\setcounter{equation}{55}
 \begin{equation}
 \label{I3}
\begin{array}{l}
  \displaystyle \mathcal{I}_3 =  \sum\limits_{l_1 = 0}^{L_1} \ \sum\limits_{l_2 = 0}^{L_2}\frac{(\pm jC)^{l_1} \ (-1)^{l_2} \ (q)^{2{l_2}}}{l_1! \ (2{l_2})!}\ \frac{\Delta {\delta ^{l_1+2{l_2}+1} _{\rho,\max }}-\Delta {\delta ^{l_1+2{l_2}+1}_{\rho,\min }}}{l_1+2{l_2}+1}.
\end{array}
\end{equation}
\end{figure*}

In Fig. \ref{newalgorithms}, the performance of the newly proposed algorithms including the alternating MLE-MLE and joint MUSIC-MLE algorithms is demonstrated. In the simulation, a BS with $M\times N=6\times 6$ antennas is used to localise and communicate with drone 2 and drone 3 without gain-phase errors. The total number of data symbols in each subframe is $T\!\!=\!\!100$, thus the maximum length of the time window is $T+1\!\!=\!\!101$. In addition, the modulation technique employed here is 16 phase-shift keying (16PSK). In each simulation, $10^5$ Monte Carlo tests are performed. In Fig. \ref{newalgorithms}, for example, MUSIC-MLE (71) refers to the result of the joint MUSIC-MLE algorithm in the time window with a length of $t+1$, where $t=70$. The result in Fig. \ref{newalgorithms} clearly indicates that both the localisation and communication performance of the two proposed algorithms become better with the increase of the length of the time window until the time window's length equals 101. Furthermore, by comparing the results of these two algorithms, it can be found that the joint MUSIC-MLE algorithm exhibits a lower performance compared to that of the alternating MLE-MLE algorithm. However, the performance gap becomes smaller with the increase of the length of the time window. By comparing the alternating MLE-MLE algorithm in JLDD and AO-ML in DLDD, both of which adopt the alternating optimisation technique, with the same simulation parameters, it can be found that both the localisation and decoding performance in alternating MLE-MLE is better than that of AO-ML, which indicates that JLDD manages to have an improved performance than DLDD. In addition, the gap widens with the increase of the time window's length. 

In Fig. \ref{newalgorithm_errors}, the result of the two proposed algorithms with gain-phase defects in the time window with a length of $101$ is provided. The simulation setup is the same as that of Fig. \ref{newalgorithms} except the latter has gain-phase defects, which are set to $\nu=1$, $ \sigma_r=0.001$ and $\tilde k=1000$ in the latter. In Fig. \ref{newalgorithm_errors}, for example, MLE-MLE-E refers to the result of the alternating MLE-MLE algorithm with gain-phase errors. The result in Fig. \ref{newalgorithm_errors} shows that gain-phase errors will affect both the localisation and communication performance of the alternating MLE-MLE and joint MUSIC-MLE algorithms. 
\section{Conclusion}\label{Sec-conc}

In this paper, we introduced the PASCAL system and evaluated its performance under gain-phase errors. Two possible PASCAL scenarios were presented. For the DLDD scenario, CRLB and average SDR were respectively derived to assess the localisation and communication performance.  For the JLDD scenario, the alternating MLE-MLE and joint MUSIC-MLE algorithms were proposed. The result indicated that gain-phase errors would not only affect the localisation performance but also the communication performance. This is because the reliability of detecting data not only depends on the channel estimation errors introduced due to gain-phase defects and noise, but is also affected by the imperfection of the
received signal itself as the received signal also contains these defects. In addition, we also found that instead of obtaining a trade-off between communication and localisation as in the conventional ISAC
system, our system model achieved a win-win situation. For the newly proposed algorithms, the result indicated that their performance continued to improve until all the symbols in each subframe are decoded as the algorithms could use more location information from the symbols in the subframe hence enhance the localisation performance, which in turn improved the decoding performance. 

\begin{appendices}
\section{ A Derivation for Theorem 1 and $\mathbb{E}[{e^{ \pm jC\Delta {\delta _{\rho}}}}]$}
To prove Theorem 1 on page \pageref{E-theorem-1}, the definition of the expectation is invoked, thus $\mathbb{E}[{\alpha^c _{\rho}}]\;\forall\alpha_{\rho}\in[0,\infty]$ can be given by
\setcounter{equation}{48}
\vspace{0.1cm}
\begin{equation}
\label{alphac}
\mathbb{E}[{\alpha^c _{\rho}}] = \int_{{0}}^{\infty} {\alpha^c _{\rho}{f}(\alpha _{\rho})d(\alpha _{\rho})},
\vspace{0.1cm}
\end{equation}
where $f({\alpha _{\rho}})$ is given in \eqref{rician PDF}.  By substituting $\small{{{I_{0} }(x)}}$ with the infinite series representation of the modified Bessel function as ${{{I_{q} }(x)}}= \sum\limits_{b = 0}^\infty  {{1}/({{b!\Gamma (b+{q} + 1)}}}) {\left({x}/{2}\right)^{2b+{q}}}$ and performing some algebraic operations, $\mathbb{E}[{\alpha _{{\rho}}^c}]$ can be written as
\begin{equation}
\label{alphac2}
\begin{array}{l}
\mathbb{E}[{\alpha^c _{{\rho}}}] = \displaystyle \frac{{e}^ { - \frac{{{\nu ^2}}}{{2{\sigma_r ^2}}}}}{{{\sigma_r ^2}}} \sum\limits_{{ b} = 0}^\infty  { \frac{{(\frac{\nu }{{2{\sigma_r ^2}}})^{2{ b}}}}{{{ b}!\Gamma ({ b} + 1)}}} {\mathcal{I}_1},
\end{array}
\end{equation}
where $ \vspace{0.1cm} \displaystyle\mathcal{I}_1=\int_0^{\infty}{\alpha _{{\rho}}^{2{ b} + c+1} {e}^{  - {{\alpha _{{\rho}}^2}}/{{2{\sigma_r^2}}}}}d(\alpha _{{\rho}})$, which can be solved by using \cite[Eq.8, pp. 346]{Integrals} as
\vspace{0.1cm}
\begin{equation}
\label{I1}
\begin{aligned}
\mathcal{I}_1= \frac{1}{2}{(\frac{1}{{2{\sigma_r ^2}}})^{  - \frac{2{ b}+c+2}{2} }}
\Gamma \left(\frac{2{ b}+c+2}{2} \right),
\end{aligned}
\vspace{0.1cm}
\end{equation}

 \begin{figure*}[ptb]
\centering 

{\includegraphics  [height=1.6in, width=7in]{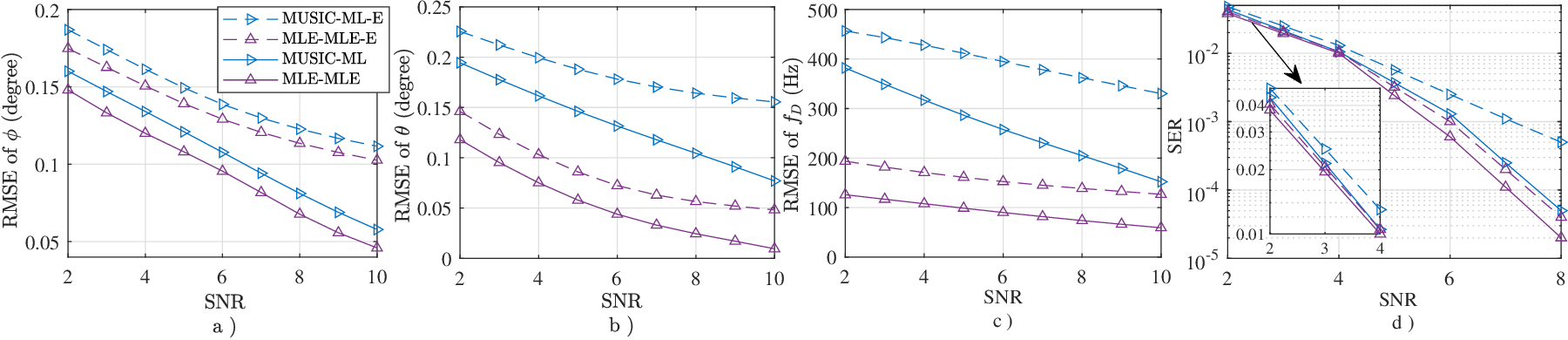}}

\caption{The effect of gain-phase errors on the alternating MLE-MLE and joint MUSIC-MLE algorithms.} 
\label{newalgorithm_errors}
\end{figure*}

By substituting the results of \eqref{I1} in \eqref{alphac2}, the closed form of $\mathbb{E}[{\alpha^c _{\rho}}]$ can be obtained, which is shown in Theorem 1.

On the other hand, $\mathbb{E}[{e^{ \pm jC\Delta {\delta _{\rho}}}}]$ can be expressed as
\setcounter{equation}{51}
\begin{equation}
\mathbb{E}[{e^{ \pm  jC\Delta {\delta _{\rho}}}}] = \int_{{\Delta {\delta _{\rho,\min }}}}^{\Delta {\delta _{\rho,\max }}} {{e^{ \pm jC\Delta {\delta _{\rho}}}}f(\Delta {\delta _{\rho}})} d\Delta {\delta _{\rho}},
\label{e59}
\end{equation}
where $f(\Delta {\delta _{\rho}})$ in \eqref{e60} can also be denoted
by using a series of Bessel functions as $ f(\Delta {\delta _{\rho}}) \!\! =\!\!\displaystyle 1/{{2\pi }}(1 + 2/{{{I_0}(\tilde k)}}$ $\sum\limits_{q = 1}^\infty  {{I_{q}}(\tilde k)} \cos [q(\Delta {\delta _{\rho}} - {\tilde \mu })])\vspace{-0.1cm}$. Then by substituting $f(\Delta {\delta _{\rho}})$ with the form of the above Bessel functions, $\mathbb{E}[{e^{\pm jC\Delta {\delta _{\rho}}}}]$ gives
\begin{equation}
\begin{array}{l}
\displaystyle \mathbb{E}[{e^{\pm jC\Delta {\delta _{\rho}}}}] =\mathcal{I}_2+ \frac{1}{{\pi {I_0}(\tilde k)}} {\sum\limits_{q = 1}^\infty  {{I_{q}}(\tilde k)}\mathcal{I}_3 }, 
\end{array}
\end{equation}
 where 
 \vspace{-0.2cm}
\begin{align}
 \mathcal{I}_2&= \displaystyle \int_{{\Delta {\delta _{\rho,\min }}}}^{\Delta {\delta _{\rho,\max }}} {\frac{{{e^{\pm jC\Delta {\delta _{\rho}}}}}}{{2\pi }} d{\Delta {\delta _{\rho}}}} \nonumber \\ \displaystyle 
&= \frac{1}{{2C\pi }}\left( \pm j{e^{\pm jC\Delta {\delta _{\rho,\min }}}} \mp  j{e^{\pm jC\Delta {\delta _{\rho,\max }}}}\right),    
\end{align}
 and 
 \begin{equation}
  \mathcal{I}_3=\displaystyle \int_{{\Delta {\delta _{\rho,\min }}}}^{\Delta {\delta _{\rho,\max }}} {{e^{\pm jC\Delta {\delta _{\rho}}}}\cos (q\Delta {\delta _{\rho}} )d{\Delta {\delta _{\rho}}}}. \ \ \ \ 
\label{e-i3}
 \end{equation}
To calculate $\mathcal{I}_3$, the 
Taylor series expansions of ${{e^{\pm jC\Delta {\delta _{\rho}}}}}$ and $ \cos (q\Delta {\delta _{\rho}} )$, which are ${{e^{\pm jC\Delta {\delta _{\rho}}}}}\!\!=\!\!\!\sum\limits_{l_1 = 0}^{L_1} {(\pm jC\Delta {\delta _{\rho}})^{l_1}}/({l_1!})\vspace{-0.2cm}$ and $\cos (q\Delta {\delta _{\rho}} )\!\!\!=\!\!\!\sum\limits_{l_2 = 0}^{L_2} {(-1)^{l_2}(q\Delta {\delta _{\rho}})^{2{l_2}}}/{(2{l_2})!}$, are employed. Then the closed-form solution is given in \eqref{I3} on the bottom of page \pageref{I3}.

\section{Complete derivations of ${\mathbb{E}}_{1}$ and ${\mathbb{E}}_{ 2}$}
\label{Appendix B}

The general expression of ${\mathbb{E}}_{1}$ and ${\mathbb{E}}_{ 2}$ can be defined as 
\setcounter{equation}{56}
\vspace{0.15cm}
\begin{equation}
\displaystyle {\mathbb{E}}_{ 3} \triangleq  {\mathbb{E}}[{{e^{j\frac{2\pi}{\lambda}  \{ \mathcal{M}d\cos {{\hat \phi }_k} + \mathcal{N}d\sin {{\hat \phi }_k}\}\sin {{\hat \theta }_k}}}}],
 \vspace{0.15cm}
\end{equation}
where $\mathcal{M} \in \{m_1 \!- \! m_2,{m_1}\! + \!{m_3} \!-\! {{ m}_2}\! -\! {{ m}_4}  \}$ and $\mathcal{N} \in \{n_1\! -\! n_2,{n_1} \!+ \!{n_3} \!-\! {{ n}_2} \!-\! {{ n}_4} \} $. ${\mathbb{E}}_{ 3}$ can be calculated using ${\mathbb{E}}_{ 3}\!\!= \!\!\displaystyle {\int_{{\hat \phi _{k,\min}}}^{{\hat \phi _{k,\max}}} \!\!\mathcal{I}_4 }f({\hat \phi _k})d{\hat \phi _k}$, where $\mathcal{I}_4$ is given by
\vspace{-0.2cm}
\begin{equation}
\label{varpi1}
\displaystyle \mathcal{I}_4={\int_{{\hat \theta _{k,\min}}}^{{\hat \theta _{k,\max}}}  {{e^{j\frac{2\pi}{\lambda} \{ \mathcal{M}d\cos {{\hat \phi }_k} + \mathcal{N}d\sin {{\hat \phi }_k}\}\sin {{\hat \theta }_k}}}} f({{\hat \theta }_k})}d{\hat \theta _k},
\end{equation}
where $f({\hat \phi _k})$ and $f({{\hat \theta }_k})$ refer to the PDF of $ \hat \phi _k$ and $\hat \theta _k$, which can be expressed as 
\begin{subequations}
    \begin{align}
        \displaystyle f({{\hat \phi }_k}) &= \frac{1}{{\sqrt {2\pi } {\sigma _\phi }}}{e^{ - \frac{1}{2}{{(\frac{{{{\hat \phi }_k} - {{ \phi }_k}}}{{{\sigma _\phi }}})}^2}}}, \\
        \displaystyle f({{\hat \theta }_k}) &= \frac{1}{{\sqrt {2\pi } {\sigma _\theta }}}{e^{ - \frac{1}{2}{{(\frac{{{{\hat \theta }_k} - {\theta _k}}}{{{\sigma _\theta }}})}^2}}}, 
    \end{align}
\end{subequations}
where ${\sigma _\phi }$ and ${\sigma _\theta }$ represent the variance of ${{\hat \phi }_k}$ and ${{\hat \theta }_k}$. Note that the means of $ \hat \phi _k$ and $\hat \theta _k$ are $ \phi _k$ and $ \theta _k$  as the localisation algorithms employed in this paper are unbiased.

By substituting the PDF of $\hat{\theta}_k$ in \eqref{varpi1} and performing some algebraic manipulations, $\mathcal{I}_4$ can be written as
\begin{equation}
\begin{array}{l}
\displaystyle \mathcal{I}_4 = \int_{{\hat \theta _{k,\min}}}^{{\hat \theta _{k,\max}}}   {C_1}{e^{j{\nu_ {\hat \phi}} \sin  {{ \hat\theta }_k} - {C_2}{{({{\hat \theta }_k} - {\theta _k})}^2}}} d{{\hat \theta }_k}, 
\end{array}\label{e-w-69}
\end{equation} 
where $\displaystyle {C_1} = {1}/({{\sqrt {2\pi } {\sigma _\theta }}})$, $\displaystyle {C_2} = {1}/({{2\sigma _\theta ^2}})$ and ${\nu_ {\hat \phi}}  = 2\pi \!/\!\lambda \{ \!\mathcal{M}d\cos {{\hat \phi }_k} + \mathcal{N}d\sin {{\hat \phi }_k}\!\}$. However, a closed-form solution for $\mathcal{I}_4$ is not feasible, and thus the small angle approximation is employed to approximate $\sin  {{ \hat\theta }_k}$ as 
$\sin  {{ \hat\theta }_k}=\sin ({{\theta }_k}+\Delta \theta_k) \approx \sin  {{ \theta }_k}+\cos  {{ \theta }_k}( {\hat { \theta }_k}-{{ \theta }_k})$. Therefore, $\mathcal{I}_4$ can be approximated as
\begin{equation}
\label{varpi2111}
\begin{array}{l}
\displaystyle \mathcal{I}_4  \!\approx\! {C_1}{e^{j{\nu_ {\hat \phi}} (\sin {{ \theta }_k}-\cos {{ \theta }_k}{ \theta }_k)}}\!\!\!\int_{{\hat \theta _{k,\min}}}^{{\hat \theta _{k,\max}}}  \!\!\!\!\!{e^{{j{\nu_ {\hat \phi}} \cos {{ \theta }_k}{\hat \theta }_k}- {C_2}{{({{\hat \theta }_k} - {\theta _k})}^2}}} \! \!d{{\hat \theta }_k}, 
\end{array}
\end{equation} 
which, by using \cite[Eq.1, pp. 108]{Integrals} and the substitution method in integral calculations, can be found as
\vspace{0.1cm}
\begin{equation}
\label{I_4}
\begin{array}{l}
\displaystyle\mathcal{I}_4= {{\frac{\sqrt{ \pi}{C_1}}{2 \sqrt{ C_2}}  ({\rm{erf}}_1-{\rm{erf}}_2){e^{j{\nu_ {\hat \phi} } \sin {\theta _k}-\frac{{\nu^2_ {\hat \phi} }\cos {\theta^2 _k}}{4{C_2}}}}}},
\end{array}
\vspace{0.1cm}
\end{equation} 
where ${\rm{erf}}_1= {\rm{erf}}(\sqrt {{C_2}} ({\theta _k}  - {\hat \theta _{k,\min}})+j{\nu_ {\hat \phi} } \cos {\theta _k}/(2\sqrt {{C_2}}))$ and ${\rm{erf}}_2= {\rm{erf}}(\sqrt {{C_2}} ({\theta _k}  - {\hat \theta _{k,\max}})+j{\nu_ {\hat \phi} } \cos {\theta _k}/(2\sqrt {{C_2}}))$. 

Afterwards, by substituting the result of $\mathcal{I}_4$ in  \label{I_4} into ${\mathbb{E}}_{ 3}$, $\displaystyle {\mathbb{E}}_{ 3}$ can be evaluated as ${\mathbb{E}}_{ 3}= {\sqrt{ \pi}{{C_1}{C_3}}/({2 \sqrt{ C_2}}})(\mathcal{I}_5-\mathcal{I}_6)$, where $\displaystyle {C_3} \!\!= \!\!{1}/{{(\sqrt {2\pi } {\sigma _\phi })}}$. $\mathcal{I}_5$ and $\mathcal{I}_6$ can be denoted using the general expression as 
\begin{equation}
\begin{array}{l}
\displaystyle \mathcal{I}_i =  \mathop \int_{{\hat \phi _{k,\min}}}^{{\hat \phi _{k,\max}}}  {\rm{erf}}_{i-4}{e^{j{\nu_ {\hat \phi} } \sin {\theta _k}-\frac{{\nu^2_ {\hat \phi} }\cos {\theta^2 _k}}{4{C_2}}- {C_4}({{\hat \phi }_k} - {\phi _k})^2}}d{{\hat \phi }_k},
\end{array}
\end{equation} 
where $\displaystyle {C_4}\! \!=\!\! {1}/({{2\sigma _\phi ^2}})$ and $i\in\{5,6\}$. By using small angle approximation for ${\nu_ {\hat \phi} }$, and then perform some algebraic manipulations, the approximated ${\nu_ {\hat \phi} }$ can be expressed as ${\nu_ {\hat \phi} }\approx C_5+{C_6} {\hat \phi}_k$, where $   C_6\!\!=\!\!2\pi/\lambda(\mathcal{N}d\cos {\phi _k}\!\!-\!\!\mathcal{M}d\sin {\phi _k})$ and
\begin{equation} C_5\!\!=\!\!\frac{2\pi}{\lambda}[\mathcal{M}d(\cos {\phi _k}+\sin {\phi _k}{\phi _k})+\mathcal{N}d(\sin {\phi _k}-\cos {\phi _k}{\phi _k})].
\end{equation}

By performing some simple algebraic manipulations, $\mathcal{I}_i$ can be simplified to
\begin{equation}
\label{varpi2}
\begin{array}{l}
 \displaystyle \mathcal{I}_i =  \mathop \int_{{\hat \phi _{k,\min}}}^{{\hat \phi _{k,\max}}} {\rm{erf}}(C_{10}+C_{11} {{\hat \phi }_k}){e^{C_7+C_8{{\hat \phi }_k}+C_9{{\hat \phi }^2_k}}}d{{\hat \phi }_k},
\end{array}
\end{equation} 
where 
\begin{subequations}
\vspace{-0.2cm}
\begin{align}
C_7\!\!&=\!\!jC_5\sin {\theta _k}\!-\!\frac{C^2_5\cos {\theta^2 _k}}{4C_2}\!-\!C_4{{ \phi }^2_k}, \\
 C_8\!\!&=\!\!jC_6\sin {\theta _k}\!-\!\frac{{C_5}{C_6}\cos {\theta^2 _k}}{2C_2}\!+\!2C_4{{ \phi }_k},
 \vspace{-0.2cm}
\end{align}  
\end{subequations}
and $C_9\!\!=\!\!-{C^2_6}\cos {\theta^2 _k}/(4C_2)-C_4 $ and $C_{11}$ is written as $C_{11}=jC_6\cos {\theta _k}/(2\sqrt{C_2})$. In addition, $C_{10}$ is given by 
\begin{equation}
    C_{10}=\sqrt{C_2}(\theta_k-\hat \theta_{k,i})+\frac{jC_5\cos {\theta _k}}{2\sqrt{C_2}},
\end{equation}
where $\theta_{k,i}=\theta_{k,\min}$ for $i=5$ and $\theta_{k,i}=\theta_{k,\max}$ for $i=6$.

Note that directly calculating the integral in \eqref{varpi2} is not feasible, thus the Taylor series of the exponential function at $a_1=C_7+C_8{{ \phi }_k}+C_9{{ \phi }^2_k}$ is utilised as
\begin{equation}
    {e^{C_7+C_8{{\hat \phi }_k}+C_9{{\hat \phi }^2_k}}}\approx \sum\limits_{l_3 = 0}^{L_3}\frac{[C_8({{\hat \phi }_k}-{{ \phi }_k})+C_9({{\hat \phi }^2_k}-{{ \phi }^2_k})]^{l3}e^{a1}}{l_3!},
\end{equation}
and the Taylor series of the error function at $a_2=C_{10}+C_{11} {{ \phi }_k}$ is also employed as
\vspace{-0.1cm}
\begin{equation}
   \!\! {\rm{erf}}(\!a_3\!)\!\approx\! {\rm{erf}}(a_2)\!+\!\sum\limits_{l_4 \!=\! 1}^{L_4}\frac{2e^{-a^2_2}(\!-\!1\!)^{l_4\!+\!1} \!H_{l_4\!-\!1}\!(a_2)(\!C_{11}({{\hat \phi }_k}\!-\!{{ \phi }_k}))^{l4}}{\sqrt{\pi}l_4!}\vspace{-0.1cm},   
\end{equation}
where $a_3=C_{10}\!+\!C_{11} {{\hat \phi }_k}$ and $H_{l_4}\!(\cdot)$ denotes the  Hermite polynomials. Note that since  ${\hat{ \phi }_k}$ is close to ${{ \phi }_k}$, the above approximations can converge very quickly and thus $L_3$ and $L_4$ are very small. Afterwards, by substituting the approximated values of the error function and exponential function and then using the binomial theorem for multiple times, the approximated $\mathcal{I}_i$ can be obtained as 
\begin{equation}
\mathcal{I}_i=\sum\limits_{l_3 = 0}^{L_3}\frac{{\rm{erf}}(a_2)e^{a_1}}{{l_3!}}\mathcal{I}_7+\sum\limits_{l_3 = 0}^{L_3}\sum\limits_{l_4 = 0}^{L_4}\frac{2e^{-a^2_2}(-C_{11})^{l_4} H_{l_4}(a_2)}{\sqrt{\pi}l_3!l_4!}\mathcal{I}_8,
\end{equation}
where 
\vspace{-0.2cm}
\begin{equation}
\mathcal{I}_7=\sum\limits_{l_5 = 0}^{l_3}\sum\limits_{l_6 = 0}^{l_5}\sum\limits_{l_7 = 0}^{l_3-l_5}C_{12}(-\phi_k)^{2{l_3}-l_5-l_6-2{l_7}}\mathcal{I}_9,
\end{equation}
 where $\small C_{12}\!\!=\!\!\left( \!\! \!{\begin{array}{*{20}{c}}
{{l_3}}\\
{{l_5}}
\end{array}}\!\! \!\right)\left( \!\!\! {\begin{array}{*{20}{c}}
{{l_5}}\\
{{l_6}}
\end{array}}\!\!\! \right)\left( \!\!\! {\begin{array}{*{20}{c}}
{{l_3\!-\!l_5}}\\
{{l_7}}
\end{array}}\!\! \!\right)C^{l_5}_8 C^{l_3\!-\!l_5}_9$ and $\mathcal{I}_9$ is 
\begin{equation}
    \mathcal{I}_9=\mathop \int_{{\hat \phi _{k,\min}}}^{{\hat \phi _{k,\max}}} {\hat \phi ^{l_6+2{l_7}}_{k}} d{{\hat \phi }_k}=\frac{{{\hat \phi^{l_6+2{l_7}+1} _{k,\max}}}-{{\hat \phi^{l_6+2{l_7}+1} _{k,\min }}}}{l_6+2{l_7}+1},
\end{equation}
and $\displaystyle \mathcal{I}_8=\sum\limits_{l_5 = 0}^{l_3}\sum\limits_{l_6 = 0}^{l_5}\sum\limits_{l_7 = 0}^{l_3-l_5}\sum\limits_{l_8 = 0}^{l_4}C_{12}(-\phi_k)^{2{l_3}+l_4-l_5-l_6-2{l_7}-l_8}\mathcal{I}_{10} \vspace{-0.2cm} $, where
\begin{equation}
    \mathcal{I}_{10}\!\!=\!\!\mathop \int_{{\hat \phi _{k,\min}}}^{{\hat \phi _{k,\max}}} \!\!{\hat \phi ^{l_6\!+\!2{l_7}\!+\!l_8}_{k}} d{{\hat \phi }_k}\!\!=\!\!\frac{{{\hat \phi^{l_6\!+\!2{l_7}\!+\!l_8\!+\!1} _{k,\max}}}\!\!-\!\!{{\hat \phi^{l_6\!+\!2{l_7}\!+\!l_8\!+\!1} _{k,\min }}}}{l_6\!+\!2{l_7}\!+\!l_8\!+\!1},
\end{equation}
\end{appendices}

\section*{REFERENCES}
\footnotesize
\def\refname{\vadjust{\vspace*{-1em}}} 

\end{document}